\documentclass[aps,twocolumn,superscriptaddress,preprintnumbers,showpacs,article,prl,nobibnotes]{revtex4-2}

\usepackage[latin9]{inputenc}
\usepackage{color}
\usepackage{verbatim}
\usepackage{amsmath}
\usepackage{amssymb}
\usepackage{graphicx}
\usepackage{esint}
\usepackage{esvect}
\usepackage{braket}
\usepackage{amsmath,bm}
\usepackage{graphicx}
\usepackage{subfigure}
\usepackage{url}
\usepackage{lipsum}
\usepackage{lipsum}
\usepackage[dvipsnames]{xcolor}
\usepackage{hyperref}
\usepackage[resetlabels]{multibib}
\newcommand{\remove}[1]{}

\newcommand{\bi}{\begin{itemize}}
\newcommand{\ei}{\end{itemize}}
\newcommand{\be}{\begin{enumerate}}
\newcommand{\ee}{\end{enumerate}}
\newenvironment{dfn}{{\vspace*{1ex} \noindent \bf Definition }}{\vspace*{1ex}}

\definecolor{OliveGreen}{rgb}{0,0.6,0}

\newcommand{\nn}{\nonumber}  %

	\newcommand{\beq}{\begin{eqnarray}}
	\newcommand{\eeq}{\end{eqnarray}}
	\newcommand{\bea}{\begin{eqnarray}\begin{aligned}}
	\newcommand{\eea}{\end{aligned}\end{eqnarray}}

\begin{document}
\title{Construction of $G_2$ symmetry in a Hubbard-type model} 
\author{Zhi-Qiang Gao}
\affiliation{Department of Physics, University of California, Berkeley, California 94720, USA}
\author{Congjun Wu}
\email{wucongjun@westlake.edu.cn}
\affiliation{New Cornerstone Science Laboratory, Department of Physics, School of Science, Westlake University, Hangzhou 310024, Zhejiang, China}
\affiliation{Institute for Theoretical Sciences, Westlake University, Hangzhou 310024, Zhejiang, China}
\affiliation{Key Laboratory for Quantum Materials of Zhejiang Province, School of Science, Westlake University, Hangzhou 310024, Zhejiang, China}
\affiliation{Institute of Natural Sciences, Westlake Institute for Advanced Study, Hangzhou 310024, Zhejiang, China}

\begin{abstract}
As the smallest exceptional Lie group and the automorphism group of the non-associative algebra octonions, $G_2$ is often employed for
describing exotic symmetry structures.
We construct $G_2$ symmetry in a self-dual Hubbard-type model with 4-component fermions in a bipartite lattice, which lies in the intersection of two $SO(7)$ algebras connected by the structure constants of octonions.
Depending on the representations of the order parameters, the $G_2$
symmetry can be spontaneously broken into either an $SU(3)$ one associated with an $S^6$ sphere Goldstone manifold, or, into $SU(2)\times U(1)$ with a Grassmannian Goldstone manifold.
In the quantum disordered states, quantum fluctuations generate the
effective $SU(3)$ and $SU(2)\times U(1)$ gauge theories for low energy fermions.
\end{abstract}
\maketitle

{\it Introduction}--- The symmetry principle is a main theme of modern physics, which distills beauty from physical phenomena. 
For example, fundamental interactions in the Standard Model are determined by gauge symmetries, including the $U(1)$ gauge theory for electromagnetism, the $SU(2) \times U(1)$ and $SU(3)$ ones for the electroweak interaction and quantum chromodynamics (QCD), respectively~\cite{weinberg2005}.
In solids, spontaneous symmetry breaking provides a guiding principle to identify phases of matter~\cite{anderson1984}.
Symmetries of high rank Lie groups, e.g., $SU(N)$, though difficult to realize in condensed matter systems, are often applied as a mathematical tool of the large-$N$ expansion to handle strong correlation effects~\cite{affleck1988,harada2003,sachdev1991}.

The development of ultra-cold atom physics opens up a realistic opportunity~\cite{wu2010,wu2012,taie2012,killian2010}.
A perspective from high symmetries (e.g., $SU(N)$ and $Sp(N)$) was proposed to study new physics of alkali and alkaline earth fermions~\cite{wu2003,wu2006}.
An exact and generic hidden symmetry of $Sp(4)$, or, isomorphically $SO(5)$ symmetry is proved in spin-$\frac{3}{2}$ systems,
%(e.g.,$^{132}$Cs, $^9$Be, $^{135}$Ba, $^{137}$Ba, $^{201}$Hg),
which can be augmented to $SU(4)$ when the interaction becomes spin-independent.
The high symmetry of $SO(5)$ without fine-tuning is rare, protecting the hidden degeneracy among collective excitations, and giving rise to the non-Abelian defects and the $SO(4)$ Cheshire charge in the quintet superfluid~\cite{wu2010a}.
%It also unifies competing orders including antiferromagnetism in different spin-tensor channels, charge-density-wave, and superconductivity~\cite{wu2006}.
The $SU(N)$ symmetry has also been widely investigated in various ultracold atom systems~\cite{gorshkov2010,cazalilla2009}, which sheds light on exploring novel phases of matter in experiments~\cite{taie2012,ozawa2018,taie2020}.

The exceptional Lie groups have attracted considerable interests in condensed matter studies~\cite{zamolodchikov1989,coldea2010,bernevig2003,
lopes2019,Lim2023}.
The $E_8$ excitations have been observed in an Ising chain antiferromagnet~\cite{coldea2010}.
Among the exceptional group family, $G_2$ is the smallest with a rank-2 and 14 dimensional algebra~\cite{agricola2008}.
It is also the automophism of the non-associative algebra of octonions~\cite{baez2002}.
Furthermore, $G_2$ is of great interests in high energy physics
since its center is trivial,
which allows to investigate whether a non-trivial center is necessary in gauge confinement~\cite{holland2003,gunaydin1996}.
A model of interacting Majorana fermions characterized by the Fibonacci topological field theory is constructed.
It is based on the $SO(7)/G_2$ coset, and exhibits exotic non-Abeliean statistics~\cite{Hu2018,luo2020,Li2023}.
The exotic physics associated with the $G_2$ group remains to be further explored, in particular, in the context of condensed matter physics with microscopic models.

In this paper, we construct a $G_2$ symmetric lattice model based on 4-component, such as spin-$\frac{3}{2}$, fermions with on-site Hubbard-type interactions. 
The $G_2$ symmetry lies in the intersection of two different $SO(7)$ symmetries: One denoted as $SO(7)_A$
below unifies the singlet superconductivity
and antiferromagnetic spin quadrupole~\cite{wu2006}, and the other one denoted as $SO(7)_B$ is related to the previous one by a self-duality between fermionic and bosonic states.
For interacting systems, if the order parameters span a 7-vector of $G_2$, the spontaneous symmetry breaking pattern is $G_2/SU(3) \cong S^6$.
If the order parameters further mix the $G_2$ adjoint representation, the residual symmetry of $SU(3)$ further splits into $SU(2)\times U(1)$, and then the Goldstone manifold becomes an oriented real Grassmannian manifold.
If the system becomes quantum disordered under strong 
fluctuations, the effective theories for low energy fermions
become the $SU(3)$, or, $SU(2)\times U(1)$ gauge theories,
both of which are crucial in high energy physics.

{\it Model Hamiltonian}--- As explained in Supplemental Material (S. M.) Sec. I~\cite{supp}, it is impossible to construct a quadratic mean-field theory with $G_2$ as the maximal residual symmetry.
Therefore the simplest way to exhibit the $G_2$ symmetry is at quartic interaction level. We define a $G_2$ invariant Hubbard-type model with spin-$\frac{3}{2}$ fermions in bipartite lattice systems in any dimension.
The free part of Hamiltonian $H_0$ is defined as
\beq
H_0 &=& -t \sum_{\left< \mathbf{ij} \right>, \sigma} \left(\psi^{\dag}_{\sigma}(\mathbf{i})
\psi_{\sigma}(\mathbf{j})+h.c. \right),
\label{eq:h0}
\eeq
where $\psi_\sigma(\mathbf{i})$ is the 4-component spinor on site $\mathbf{i}$
with $\sigma=\pm \frac{3}{2}$, $\pm \frac{1}{2}$ representing the
spin components;
$\langle \mathbf{ij} \rangle$ represents the nearest neighbor
bonding between two sublattices;
the chemical potential is set to zero to ensure the particle-hole symmetry.
The interaction part 
$H^{G_2}_\mathrm{int}$ is 
\beq
H^{G_2}_\mathrm{int}&=&u\sum_{\mathbf{i}}
M_{ab}(\mathbf{i})M_{ab}(\mathbf{i}) %\nonumber \\
+ v \sum_{\mathbf{i}} 
M^\prime_{ab}(\mathbf{i})M^\prime_{ab}(\mathbf{i}),~~
\label{eq:g2}
\eeq
where $u$ and $v$ are interaction parameters.
$M_{ab}$ with $0\le a, b\le 6$ span the $SO(7)_A$ algebra, which are antisymmetric under exchanging $a$ and $b$, and $M^\prime_{ab}$ is the dual of $M_{ab}$ spaning the $SO(7)_B$ algebra. Both will be explained below.

The 4-component spinor 
$\psi_\sigma(\mathbf{i})$ spans a local $SO(8)$
algebra with generators denoted as $M_{ab}(\mathbf{i})$ ($0\le a, b\le 7$) below.
Its $SO(5)$ subalgebra with $1\le a,b\le 5$
is isomorphic to $Sp(4)$ defined as $M_{ab}(\mathbf{i})=-\frac{1}{2}
\psi^\dagger(\mathbf{i}) \Gamma^{ab} \psi(\mathbf{i})$.
Here $\Gamma^{ab}=-\frac{i}{2}[\Gamma^a, \Gamma^b]$,
and $\Gamma^a$'s are the four by four gamma matrices anticommuting with each other, spanning a rank-2 Clifford algebra.
The $SO(5)$ subalgebra is extended to $SO(7)_A$ by including
$M_{06}(\mathbf{i})=(n(\mathbf{i})-2)/2$ with
$n(\mathbf{i})=\psi^\dagger(\mathbf{i}) \psi(\mathbf{i})$
the particle number, and other generators are quintet Cooper pair defined as $M_{0a}(\mathbf{i})+iM_{a6}(\mathbf{i})=-\frac{i}{2}
\text{sgn}(\mathbf{i})\psi^\dagger (\mathbf{i}) \Gamma^a R \psi^\dagger(\mathbf{i})$,
where $\text{sgn}(\mathbf{i})=\pm 1$ alternates on two sublattices, and $R=\Gamma_1\Gamma_3$ is the charge conjugation matrix~\cite{wu2006}.
Other generators transform as an 
$SO(7)_A$ 7-vector, organized as $V_a(\mathbf{i})=M_{a7}(\mathbf{i})$ ($0\le a\le 6$).
Concretely,
$V_0(\mathbf{i})-iV_6(\mathbf{i})=\frac{1}{2}\mbox{sgn}(\mathbf{i}) \psi^\dagger(\mathbf{i}) R\psi^\dagger(\mathbf{i})$
is the singlet Cooper pairing operator,
and $V_a(\mathbf{i})=\frac{1}{2}\psi^\dagger(\mathbf{i}) \Gamma^a \psi(\mathbf{i})$
with $1\le a \le 5$ are spin qudrupole operators.
A detailed explanation of the $SO(8)$ algebra is presented
in the S. M. Sec. II~\cite{supp}.

Eq.~(\ref{eq:g2}) in the case of $v=0$ is a special case of the generic spin-$\frac{3}{2}$ Hubbard model~\cite{wu2003}, which has only two interaction channels with strengths denoted as $U_{0,2}$, respectively.
The $Sp(4)$ symmetry was proved for general values of $U_{0,2}$.
Eq.~(\ref{eq:g2}) with $v=0$ corresponds to the condition of $u=\frac{2}{3}U_2=-2U_0$ combined with the particle-hole symmetry, {\it i.e.}, half-filling on a bipartite lattice~\cite{wu2003}.
The global $SO(7)_A$ generators are defined as $M_{ab}=\sum_i M_{ab}(i)$ for $0\le a,b\le 6$, which commutes with $H_0$.
Hence, Eq.~(\ref{eq:g2}) is $SO(7)_A$ invariant at $v=0$.

{\it $G_2$ symmetry}---
Octonions are the non-associative division algebra defined as
$q=q_0+\sum_{i=1}^7 q_i \hat e_{i-1}$, where
$q_{0\sim 7}$ are real numbers;
$\hat e_i$ are imaginary units anticommuting with each other
satisfying $\hat e_i \hat e_j=-\delta_{ij} + C_{ijk} \hat e_k$.
The structure constant $C_{ijk}$ is a fully antisymmetric tensor, satisfying
$C_{ijk} C_{lmk}=\delta_{il} \delta_{jm}-\delta_{im}\delta_{jl} +
C_{ijlm}$ and $C_{ijlm}$ is the dual of $C_{ijk}$ as
$C_{ijlm}=\frac{1}{6}\epsilon_{ijlmpqk} C_{pqk}$ with $\epsilon_{ijlmpqk}$ the Levi-Civita symbol in 7 dimensions~\cite{baez2002,gunaydin1996}.
Any choice of the octonion structure constant $C_{ijk}$ 
yields a $G_2$ invariant Hamiltonian.
Nevertheless, for later convenience, we choose that 
$$
C_{031} = C_{052} = C_{064} = C_{126} = C_{154} = C_{243} = C_{365} = 1.
$$
Details of $C_{ijk}$ and its dual $C_{ijlm}$ are given in S. M. Sec. III~\cite{supp}. 

The $G_2$ generators $G_{ab}$ can be extracted from $SO(7)_A$ via the octonion structure constant as,
\beq
M_{ab}(\mathbf{i})=G_{ab}(\mathbf{i}) + \frac{1}{\sqrt 3} C_{abc} T_c(\mathbf{i}), 
~~ 0\le a, b, c \le 6.
\eeq
7 constraints that $C_{abc}G_{bc}=0$ are satisfied by $G_{ab}$,
and hence, only 14 of them are independent~\cite{gunaydin1996}. $T_c$ belongs to the coset $SO(7)_A/G_2$ and transforms as a 7-vector of $G_2$. Note that the $SO(7)_A$ vector $V_a$ is also a $G_2$ vector.
$G_{ab}$ and $T_c$ are extracted from $M_{ab}$ as $G_{ab}(\mathbf{i})=\frac{2}{3}  M_{ab}(\mathbf{i}) +\frac{1}{6} C_{abcd}
M_{cd}(\mathbf{i})$, and $T_a(\mathbf{i})=\frac{1}{2\sqrt{3}} C_{abc}  M_{bc}(\mathbf{i})$.
A remarkable property is that a new $SO(7)_B$ algebra can be constructed, sharing the same $G_2$ subalgebra, as
$M^\prime_{ab}(\mathbf{i})=G_{ab}(\mathbf{i}) +
\frac{1}{\sqrt 3} C_{abc} T^\prime_c(\mathbf{i})$,
and the coset $SO(8)/SO(7)_B$ is spanned by the 7-vector $V_a^\prime$.
The two versions of $G_2$ vectors 
based on $SO(7)_B$ and $SO(7)_A$
are transformed into each other by an ``$120^\circ$ rotation" in the $T$-$V$ plane as
$T_a^\prime+i V_a^\prime= e^{i\frac{2}{3}\pi} (T_a+iV_a)$~\footnote{An immediately question is what if the rotation in $T$-$V$ plane is $-120^\circ$, instead of $120^\circ$. In fact, if we define $T_a^{\prime\prime}+i V_a^{\prime\prime}= e^{-i\frac{2}{3}\pi} (T_a+iV_a)$, and accordingly define $M^{\prime\prime}_{ab}$, it will indeed give the third $SO(7)$ symmetry! This third $SO(7)$ symmetry is self-dual under $\chi_0(\mathbf{i})$, namely $\chi_0(\mathbf{i}) M^{\prime\prime}_{ab}(\mathbf{i})\chi_0(\mathbf{i})=M^{\prime\prime}_{ab}(\mathbf{i})$ and $\chi_0(\mathbf{i}) V^{\prime\prime}_{a}(\mathbf{i})\chi_0(\mathbf{i})=-V^{\prime\prime}_{a}(\mathbf{i})$. However, a straightforward algebra shows its Casimir equal to a $c$-number, 21, instead of some operator in terms of $\psi_\sigma(\mathbf{i})$. Therefore, this third $SO(7)$ symmetry is absent in the lattice model. Nevertheless, in a sequential work~\cite{Gao2024} we show that it can emerge in the continuum limit, and fulfill the triality structure of $SO(8)$ together with $SO(7)_A$ and $SO(7)_B$.}.

Now Eq.~(\ref{eq:g2}) is explicitly $G_2$ symmetric.
The global $SO(7)_B$ generators are defined via $M^\prime_{ab}=
\sum_\mathbf{i} M_{ab}^\prime(\mathbf{i})$, which commute with both $H_0$ plus the
$v$-term in $H^{G_2}_\text{{int}}$.
When both $u$ and $v$ are non-zero, only the $G_2$ symmetry exists as
the intersection between $SO(7)_A$ and $SO(7)_B$. 
The union of $SO(7)_B$ and its vector $V_a^\prime$ is the same $SO(8)$
algebra spanned by $SO(7)_A$ and $V_a$, {\it i.e.},
$SO(8)=SO(7)_A\bigcup V=SO(7)_B\bigcup V^\prime$.
The Casimirs of $SO(7)_{A,B}$ are related via the octonion structure constants as
$C_A(\mathbf{i})=M_{ab}(\mathbf{i}) M_{ab} (\mathbf{i}) =\frac{1}{4}C_{abcd}M^\prime_{ab}(\mathbf{i}) M^\prime_{cd}(\mathbf{i})$, and vice versa $C_B(\mathbf{i})=M^\prime_{ab}(\mathbf{i}) M^\prime_{ab} (\mathbf{i}) =\frac{1}{4}
C_{abcd}M_{ab}(\mathbf{i})M_{cd}(\mathbf{i})$.

The Cartan subalgebra of $G_2$ can be defined as
$H_1=\frac{\sqrt 3}{2}\sum_\mathbf{i} G_{15}(\mathbf{i})+G_{23}(\mathbf{i})$ and $H_2=\frac{3}{2}\sum_\mathbf{i} G_{06}(\mathbf{i})$.
A reorganization of $H_{1,2}$ gives rise to two commutable conserved $U(1)$ quantities,
\beq
Q_1=\sum_\mathbf{i} n_{-\frac{1}{2}}(\mathbf{i})+n_{-\frac{3}{2}}(\mathbf{i}), ~
Q_2=\sum_\mathbf{i} n_{\frac{1}{2}}(\mathbf{i})-n_{-\frac{1}{2}}(\mathbf{i}).
~
\eeq
However, the total particle number $M_{06}$ (up to a constant) is no longer conserved.
Details of the $G_2$ and $SO(7)_B$ algebras are given in S. M. Sec. III~\cite{supp}.

{\it Self-duality}---
The above two $SO(7)$ symmetries are dual to each other via the Majorana operator 
\beq
\chi_0(\mathbf{i})=e^{-i\text{sgn}(\mathbf{i}) \pi/4} \psi_\frac{3}{2}(\mathbf{i})+e^{i\text{sgn}(\mathbf{i}) \pi/4}\psi^\dagger_\frac{3}{2}(\mathbf{i}),
\eeq
which is an $G_2$ singlet, or, scalar. 
The duality is revealed in the relations of $M^\prime_{ab}(\mathbf{i})=\chi_0(\mathbf{i})M_{ab}(\mathbf{i})\chi_0(\mathbf{i})$,
as well as 7-vectors $T^\prime_{a}(\mathbf{i})=\chi_0(\mathbf{i})T_{a}(\mathbf{i})\chi_0(\mathbf{i})$ and $V^\prime_{a}(\mathbf{i})=-\chi_0(\mathbf{i})V_{a}(\mathbf{i})\chi_0(\mathbf{i})$.
This implies a self-duality of the Hamiltonian Eq.~(\ref{eq:g2}). 
Define $\mathbb{Z}_2$ operator $P=i^\mathcal{N}\prod_\mathbf{i}\chi_0(\mathbf{i})$ satisfying $P^2=P^\dag P=1$ with $\mathcal{N}$ the total site number. 
It can be shown that $PH_0P=H_0$, and $PH^{G_2}_\text{int}(u,v)P=H^{G_2}_\text{int}(v,u)$, proving the self-duality. 
When $u=v$, this self-duality is promoted to a $\mathbb{Z}_2$ symmetry, which protects the gaplessness of the Majorana mode $\chi_0(\mathbf{i})$.

Consider the solution to a single site problem based on $H_{\text{int}}^{G_2}$.
The explicit expression of $H^{G_2}_\text{{int}}$ in terms of fermion operators is presented in S. M. Sec. IV~\cite{supp}.
The onsite Hilbert space is 16 dimensional,
including 8 bosonic and 8 fermionic states with even and odd fermion parity, respectively.
7 states of the bosonic sector are denoted as B$_7$, 
composed of the empty state $|0\rangle$, the fully occupied one $|F\rangle=\psi_{\frac{3}{2}}^\dag\psi_{\frac{1}{2}}^
\dag\psi_{-\frac{1}{2}}^\dag\psi_{-\frac{3}{2}}^\dag\left| 0\right>$,
and the spin quintet ones $|q_a\rangle =-\frac{i}{2}\psi^\dagger \Gamma^aR\psi^\dagger|0\rangle$. They form a 7-dimensional vector representation of the $G_2$ group. The rest one is the 2-fermion spin singlet state $|s\rangle=\frac{1}{2}\psi^\dagger R \psi^\dagger |0\rangle$ denoted as B$_1$, which is a singlet of $G_2$.
The 8 fermionic states also split into a 7-vector and a singlet
representations of $G_2$, denoted as F$_7$ and F$_1$, respectively.

The duality manifests via the one-to-one correspondence between the onsite fermionic and bosonic states in the onsite Hilbert space. 
As summarized in Table~\ref{table1},
states of B$_7$ and F$_7$, and those of B$_1$ and F$_1$ can be organized as pairs sharing the same $G_2$ weights, which can be transformed into each other by applying  the Majorana operator $\chi_0(\mathbf{i})$.
Interestingly, B$_1$ and B$_7$ remain irreducible for
$SO(7)_A$, while F$_1$ and F$_7$ together form the 8-dimensional spinor representation of $SO(7)_A$.
In contrast, F$_1$ and F$_7$ remain irreducible for $SO(7)_B$,
while B$_1$ and B$_7$ together span the spinor representation
of $SO(7)_B$. 

%%%%%%%%%%%%%%%%%%%%%%%%%%%%%%%%%%%%%%%%%%%%%%%
\begin{table}[!htbp]
\begin{tabular}{|c|c|c|}
\hline
Weight $\mu$ & Bosonic (B$_7$) & Fermionic (F$_7$) \\
\hline
$(\frac{\sqrt{3}}{2},\frac{1}{2})$ &
$\frac{1}{\sqrt{2}}(|q_2 \rangle  -i|q_3\rangle)$& $\psi_{\frac{3}{2}}^\dag\psi_{\frac{1}{2}}^\dag\psi_{-\frac{3}{2}}^\dag\left| 0\right>$\\
\hline
$(0,1)$ & $|F\rangle$ & $\psi_{\frac{1}{2}}^\dag\psi_{-\frac{1}{2}}^\dag\psi_{-\frac{3}{2}}^\dag\left| 0\right>$\\
\hline
$(-\frac{\sqrt{3}}{2},\frac{1}{2})$ & $\frac{1}{\sqrt 2}(|q_1\rangle+i|q_5\rangle) $& $\psi_{\frac{3}{2}}^\dag\psi_{-\frac{1}{2}}^\dag\psi_{-\frac{3}{2}}^\dag\left| 0\right>$\\
\hline
$(-\frac{\sqrt{3}}{2},-\frac{1}{2})$ & $\frac{1}{\sqrt 2}(|q_2\rangle+i|q_3\rangle) $ & $\psi_{-\frac{1}{2}}^\dag\left| 0\right>$\\
\hline
$(0,-1)$ & $| 0\rangle$ & $\psi_{\frac{3}{2}}^\dag\left| 0\right>$ \\
\hline
$(\frac{\sqrt{3}}{2},-\frac{1}{2})$ & $\frac{1}{\sqrt 2}(|q_1\rangle-i|q_5\rangle) $
& $\psi_{\frac{1}{2}}^\dag\left| 0\right>$\\
\hline
$(0,0)$ & $|q_4\rangle$ &  $\frac{1}{\sqrt{2}}(\psi_{-\frac{3}{2}}^\dag+\text{sgn}(\mathbf{i})i\psi_{\frac{3}{2}}^\dag\psi_{\frac{1}{2}}^\dag\psi_{-\frac{1}{2}}^\dag)\left| 0\right>$  \\
\hline
\hline
Weight $\mu$ & Bosonic (B$_1$) & Fermionic (F$_1$)  \\
\hline
(0,0)&$|s\rangle$& $\frac{1}{\sqrt{2}}(\psi_{-\frac{3}{2}}^\dag-\text{sgn}(\mathbf{i})i\psi_{\frac{3}{2}}^\dag\psi_{\frac{1}{2}}^\dag\psi_{-\frac{1}{2}}^\dag)\left| 0\right>$
 \\
\hline
\end{tabular}
%\centering
\caption{Classification of the onsite states in terms of the $G_2$ weights.
The B$_7$ sector includes the empty state, the fully occupied state, and the spin quintet states.
The B$_1$ state is a two-fermion spin singlet
state. 
The pair of bosonic and fermionic states can be transformed into each other by applying $\chi_0(\mathbf{i})$.
}
\label{table1}
\end{table}
%%%%%%%%%%%%%%%%%%%%%%%%%%%%%%%%%%%%%%%%%%%

\begin{figure}[htbp]
\centering
\includegraphics[width=0.7\linewidth]{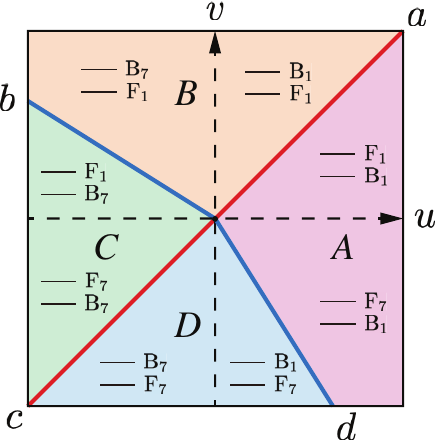}
\caption{The energy levels of a single site problem.
The lowest two energy levels are illustrated in each region.
The $SO(7)_{A,B}$ symmetries are restored along
the $u=0$ and $v=0$ lines, respectively.
The self-duality is manifested in the
reflection symmetry of the diagram
with respect to the $a$-$c$ line with $u=v$.
The $b$ and $d$ lines have $u=-7v$ and $v=-7u$, respectively.}
\label{fig:singlesite}
\end{figure}

The energy level diagram of the single site problem in terms of $u$ and $v$ is shown in Fig.~[\ref{fig:singlesite}].
The B$_1$, F$_1$, B$_7$, and F$_7$ states are the lowest energy states in the regions of $A$, $B$, $C$, and $D$, respectively,
with their energies calculated as $E_{\text{B}_1}=0$, $E_{\text{F}_1}=\frac{21}{2}(u-v)$, $E_{\text{B}_7}=12u$, and $E_{\text{F}_7}=\frac{21}{2} u+\frac{3}{2}v$.
Along the lines of $u=0$ and $v=0$, the $SO(7)_A$ and $SO(7)_B$ symmetries are restored, respectively.
The degeneracy patterns of the lowest energy levels are unaffected,
while the second lowest energy levels becomes 8-fold degenerate:
The 8 fermionic (bosonic) states form the spinor representation
of $SO(7)_{A(B)}$.
Along lines $a$ and $c$ where $u=v$, the bosonic and fermionic
states with the same $G_2$ representations become degenerates.
In contrast, along line $b$, the bosonic states B$_7$ become degenerate with the fermionic one F$_1$, and a similar degeneracy occurs along line $d$.
These features may signature possible supersymmetry structures which will be deferred for future studies~\cite{Gao2024}.

%%%%%%%%%%%%%%%%%%%%%%%%%%%%%%%%%%%%%%%%%%%
%%%%%%%%%%%%%%%%%%%%%%%%%%%%%%%%%%%%%%

{\it Spontaneous symmetry breaking}--- 
Next we consider the symmetry breaking patterns of the Hamiltonian Eqs.~(\ref{eq:h0}) and (\ref{eq:g2}), as summarized in Table~\ref{tab:symm}, which occur for sufficiently strong $u$ or $v$.
We define order parameters $O^G_{ab}=\sum_\mathbf{i} \text{sgn}(\mathbf{i}) G_{ab}(\mathbf{i})$,
$O^T_{a}=\sum_\mathbf{i} \text{sgn}(\mathbf{i}) T_a(\mathbf{i})$, and $O^V_a=\sum_\mathbf{i} \text{sgn}(\mathbf{i}) V_a(\mathbf{i})$.
The sign convection to these term is that their particle-hole part exhibits staggered signs on different sublattices, while the particle-particle part has a uniform sign.
These order parameters transform as $G_2$'s adjoint, vector, and vector representations, respectively.

In regions $A$ and $B$ of Fig.~[\ref{fig:singlesite}], order parameters connect the lowest energy $G_2$ singlets to the next $G_2$ vector states under the same fermion parity, forming a $G_2$ vector.
Since $O^T_a$ and $O^V_a$ are both $G_2$ vectors, they generally should mix. 
The action of $G_2$ is transitive over the $S^6$ sphere~\cite{agricola2018}. Therefore the order parameter can be generally represented as
$O(\mathbf{i})=\hat h_a (\cos\theta O^V_a(\mathbf{i})- \sin\theta O^T_a(\mathbf{i}))$, where
$\hat h$ is a unit vector on $S^6$, and $\theta$ is the mixing angle.
Along the $SO(7)_A$ line, $\theta=0$, and the order
is the $SO(7)_A$ vector $O^V_a(\mathbf{i})$.
Similarly, $\theta=\frac{\pi}{3}$ along the SO(7)$_B$ line.
The mean-field decomposition of the $G_2$ interaction becomes (see S. M. Sec. V for details of mean-field decompositions~\cite{supp})
\beq
H_\mathrm{MF}^{A,B}=2\sqrt{3}\Delta \sum_\mathbf{i}  \text{sgn}(\mathbf{i}) \hat h_a (\cos\theta V_a(\mathbf{i})-
\sin\theta T_a(\mathbf{i})),
\eeq
where $\Delta$ is the magnitude of the order parameter carrying the energy unit.
In the 8-component Nambu spinor representation
$\Psi(\mathbf{i}) =(\psi_\sigma(\mathbf{i}), \psi^\dagger_\sigma(\mathbf{i}))^\mathbf{T}$,
where $8=4$ spin $\times$ 2 particle-hole, the energy degeneracy pattern becomes $(3\oplus 1)^2$.
The 2-fold redundancy is due to particle-hole symmetry.
The fermion excitations are generally speaking gapped as
$E_t=\pm (\epsilon_\mathbf{k}^2+\Delta^2 (\sqrt 3 \cos\theta +\sin\theta)^2)^{\frac{1}{2}}$ for the triplets
and $E_s=\pm (\epsilon_\mathbf{k}^2+3\Delta^2(\cos\theta-\sqrt 3\sin\theta)^2)^{\frac{1}{2}}$ for the singlets,
respectively, where $\epsilon_\mathbf{k}$ is the band dispersion.
If we take $\hat h_a$ along the 4th direction, then $O^T_4$ and $O^V_4$ are diagonal in Nambu basis.
The unbroken symmetry is $SU(3)$ associated with the fundamental 3-spinor spanned by $\psi_{SU(3)}=(\psi_\frac{1}{2},\psi_{-\frac{1}{2}}, \psi^\dagger_{-\frac{3}{2}})^\mathbf{T}$, and its generators are given in S. M. Sec. V~\cite{supp}. The remaining fermion $\psi^\dagger_{-\frac{3}{2}}$ is a singlet under $SU(3)$.

The above spontaneous symmetry breaking results in the Goldstone manifold $G_2/SU(3) \cong S^6$.
Considering quantum fluctuations 
of $\hat h (\mathbf{i},t)$, and
denote $U^\dagger (\mathbf{i},t) O(\mathbf{i},t) U(\mathbf{i},t)= O_4$ where $U(\mathbf{i},t)$
is the $G_2$ operation in the 8-component Nambu basis.
The low energy effective theory is composed by an $SU(3)$ gauge theory of 3-component fermions and an $SU(3)$-neutral fermion. In the continuum representation, the gauge field is defined via the non-Abealian Berry phase,
\beq
A_\mu (\mathbf{x},t) =\mathcal{P} \left(U(\mathbf{x},t)^\dagger \partial_\mu U(\mathbf{x},t)\right)\mathcal{P},
\label{eq:gauge}
\eeq
where $\mathcal{P}$ is the projection operator into the 3-dimensional subspace.

Along the $SO(7)_A$ line with $u>0$, the order parameters $O^V_a$ include the singlet pairing and the antiferromagnetic spin quadrupole orders~\cite{wu2003}.
Although the symmetry breaking pattern is different, the Goldstone
manifold remains $S^6\cong SO(7)/SO(6) \cong \mathrm{Spin(7)}/SU(4)$.
If $\hat h$ is along the 4th direction, the unbroken $SU(4)_A$
symmetry is associated with the 
$\psi_{SU(4)_A}=
(\psi_\frac{1}{2},\psi_{-\frac{1}{2}},
\psi^\dagger_\frac{3}{2}, \psi^\dagger_{-\frac{3}{2}})^\mathbf{T}$.
Similarly, along the $SO(7)_B$ line
with $v>0$, the unbroken $SU(4)_B$ spinor is spanned by
$\psi_{SU(4)_B}=(\psi_\frac{3}{2},\psi_\frac{1}{2},\psi_{-\frac{1}{2}}, \psi^\dagger_{-\frac{3}{2}})^\mathbf{T}$.
Considering quantum fluctuations, the $SU(4)$ gauge theory for 4-component fermions is arrived.

%%%%%%%%%%%%%%%%%%%%%%%%%%%%%%%%%%
\begin{table}[!htbp]
\begin{tabular}{|c|c|c|}
\hline
Region & Order parameter & Ground state manifold \\
\hline
$A,B$ & $G_2$ vector rep. & $G_2/SU(3)\cong S^6$ \\
\hline
$C,D$ & $G_2$ adjoint rep. & $G_2$/$[SU(2)\times U(1)] \cong
\mbox{Gr}_5^+(\mathbb{R}^7)$\\
\hline
\end{tabular}
\centering
\caption{The symmetry breaking patterns and the configurations of the Goldstone manifold.
The unbroken symmetries are enlarged to $SO(7)_A$ along the $u$-axis, and $SO(7)_B$ along the $v$-axis.
Although the symmetry breaking patterns are different,
the Goldstone manifolds remain $S^6$ or
$\mbox{Gr}_5^+(\mathbb{R}^7)$.
}
\label{tab:symm}
\end{table}
%%%%%%%%%%%%%%%%%%%%%%%%%%%%%%%%

Now we consider regions $C$ and $D$ in which the lowest onsite energy levels are degenerate $G_2$ 7-vectors.
Hence, the order paramaters could connect different states among them, including all the possibilities of $O^G$, $O^T$ and $O^V$.
Consider a mixture of $O^G_{06}$, $O^T_4$ and $O^V_4$, which remain
diagonal in the Nambu basis.
Then the mean-field Hamiltonian becomes
\beq
&& H_\mathrm{MF}^{C,D}=\Delta \sum_\mathbf{i}  \text{sgn}(\mathbf{i}) \Big (\sin\phi \big( \cos\theta V_4(\mathbf{i})-
\sin\theta T_4(\mathbf{i}) \big) \nn \\
&&\qquad\qquad\qquad\qquad\quad~ +\cos\phi G_{06}(\mathbf{i}) \Big).
\label{eq:su2mf}
\eeq
As for the fermion excitation spectra, the degeneracy pattern becomes
$(2\oplus 1\oplus 1)^2$.
The 3-fold degeneracy in the regions $A$ and $B$ further splits into $2\oplus 1$ due to the mixing of the order parameter in $G_2$ adjoint representation, and again the repeated degeneracy is also attributed to particle-hole symmetry. 
The residual symmetry is consequently $SU(2)\times U(1)$ broken from $SU(3)$.
Correspondingly, the $SU(3)$ fundamental spinor splits into an $SU(2)$ fundamental spinor $\psi_{SU(2)}=(\psi_\frac{1}{2},\psi_{-\frac{1}{2}})^\mathbf{T}$ and a $U(1)$ spinor $\psi^\dagger_{-\frac{3}{2}}$.
A general order parameter configuration
is obtained by applying a space-time dependent $G_2$ rotation
$U(\mathbf{i},t)$ on that of Eq.~(\ref{eq:su2mf}), and an $SU(2)\times U(1)$ gauge
field can be defined in parallel to Eq.~(\ref{eq:gauge}).
The Goldstone manifold is a real oriented Grassmannian manifold $G_2/[SU(2)\times U(1)] \cong \mbox{Gr}_5^+(\mathbb{R}^7)$.
Its Homotopy group $\pi_2(\mbox{Gr}_5^+(\mathbb{R}^7))\cong \mathbb{Z}$ is non-trivial, with all higher homotopy groups trivial.
This implies that in low dimensions, topological defects characterized by $\mathbb{Z}$ can emerge in the (quasi-)long range order.

Along the $SO(7)_{A}$ line with $u<0$ and the $SO(7)_B$ line with $v<0$, 
the order parameters should within the adjoint representations of $SO(7)_A$ and $SO(7)_B$, respectively.
Without loss of generality, $\sum_\mathbf{i} \text{sgn}(\mathbf{i}) M_{06}(\mathbf{i})$
and $\sum_\mathbf{i} \text{sgn}(\mathbf{i}) M_{06}^\prime (\mathbf{i})$ are taken 
as the order paramters for the former and
the latter cases, respectively. %which correspond to $\theta=\frac{\pi}{2}$, $\phi=\frac{\pi}{6}$, and $\theta=\frac{\pi}{6}$, $\phi=-\frac{\pi}{6}$, respectively. 
In both cases, the degeneracy patterns change to $4\oplus 4$, reflecting the symmetry breaking pattern from $SO(7)$ into
$Sp(4)\times U(1)$.
The $Sp(4)$ spinors are also 4-dimensional
as $\psi_{Sp(4)_A}=(\psi_\frac{3}{2},\psi_\frac{1}{2},\psi_{-\frac{1}{2}},\psi_{-\frac{3}{2}})^\mathbf{T}$ and $\psi_{Sp(4)_B}=(\psi^\dagger_\frac{3}{2},\psi_\frac{1}{2},\psi_{-\frac{1}{2}}, \psi_{-\frac{3}{2}})^\mathbf{T}$
along $SO(7)_A$ and $SO(7)_B$ symmetric lines, respectively.
The Goldstone manifold becomes Spin$(7)/[Sp(4)\times U(1)]$, which remains $\mbox{Gr}_5^+(\mathbb{R}^7)$.
Again, when considering quantum fluctuations, the effective theory for
fermions becomes the $Sp(4)$ gauge theory.

The symmetry breaking patterns and possible phase transitions along the region boundaries $a$, $b$, $c$ and $d$ are even more involved.
For two phases separated by $b$ and $d$ lines, the symmetry group of one phase is the subgroup of the other, {\it i.e.}, $SU(2)\times U(1) \subset SU(3)$. 
Therefore, the phase transitions should have a spontaneous breaking nature. 
As for lines $a$ and $c$, the phases separated by them are with the same symmetry. On these two lines, due to the emergent $\mathbb{Z}_2$ symmetry promoted from the self-duality of the Hamiltonian, and its protecting gapless Majorana mode $\chi_0$, we conjecture that they are still phase transition lines. Possible supersymmetry structures, which is shown to indeed emerge in (1+1)D~\footnote{In the sequential work~\cite{Gao2024} we show that in (1+1)D spacetime supersymmetry can emerge as a tricritical Ising conformal field theory with central charge $c=\frac{7}{10}$. In fact, our model with $u=v$ reproduces the (lattice version of) Fibonacci topological field theory constructed by 7 interacting Majorana fermions~\cite{Hu2018}, while the latter one is shown to indeed have central charge $c=\frac{7}{10}$.}, may play crucial roles along these lines~\cite{Grover2014}. The possible phase transitions and supersymmetries in higher dimensions are deferred for future studies. Meanwhile, in the sequential work~\cite{Gao2024} we show, both analytically and numerically, that in (1+1)D the phase boundaries $a$ and $c$ lines are indeed located at $u=v$, and $b$ and $d$ lines are located at $u=-7v$ and $v=-7u$, respectively. These results corroborate our on-site and mean-field analysis of the phase boundaries.

{\it Conclusions}---
In summary, we have constructed a lattice 4-component fermion model with an explicit $G_2$ symmetry, which, to our knowledge, is the first $G_2$ model with microscopic lattice interaction.
This exceptional symmetry is a common subgroup of two different $SO(7)$ symmetries, both of which lie in the $SO(8)$ algebra and are related by the octonion structure constants via a self-duality mapping connecting bosonic and fermionic states.
The spontaneous symmetry breaking patterns are $G_2/SU(3)\cong S^6$, or, $G_2/[SU(2)\times U(1)]\cong \mbox{Gr}_5^+(\mathbb{R}^7)$~\footnote{For $G_2$, another kind of important symmetry breaking pattern is $G_2$ broken to $SU(2)\times SU(2)$. However, in our model, this symmetry breaking pattern does not appear naturally.}.
When quantum fluctuations dominate leading to disordered ground states, the low energy fermions are governed by the non-Abeliean gauge theories of the types of $SU(3)$ and $SU(2)\times U(1)$.
The corresponding gauge theory structures are similar to those of the strong and electroweak interactions in high energy physics.

\textit{Acknowledgements}.
ZQG and CW acknowledge F. Wang, B. Chen, H. Yang, Y.-Q. Wang, C. Liu, V. Calvera, Y. Tan, J.-C. Feng, and Y. P. Wang for helpful discussions.
CW is supported by the National Natural Science Foundation of China under the Grant No. 12234016 and No. 12174317.
This work has been supported by the New Cornerstone Science Foundation.

\bibliography{spin32}

\begin{thebibliography}{35}%
\makeatletter
\providecommand \@ifxundefined [1]{%
 \@ifx{#1\undefined}
}%
\providecommand \@ifnum [1]{%
 \ifnum #1\expandafter \@firstoftwo
 \else \expandafter \@secondoftwo
 \fi
}%
\providecommand \@ifx [1]{%
 \ifx #1\expandafter \@firstoftwo
 \else \expandafter \@secondoftwo
 \fi
}%
\providecommand \natexlab [1]{#1}%
\providecommand \enquote  [1]{``#1''}%
\providecommand \bibnamefont  [1]{#1}%
\providecommand \bibfnamefont [1]{#1}%
\providecommand \citenamefont [1]{#1}%
\providecommand \href@noop [0]{\@secondoftwo}%
\providecommand \href [0]{\begingroup \@sanitize@url \@href}%
\providecommand \@href[1]{\@@startlink{#1}\@@href}%
\providecommand \@@href[1]{\endgroup#1\@@endlink}%
\providecommand \@sanitize@url [0]{\catcode `\\12\catcode `\$12\catcode
  `\&12\catcode `\#12\catcode `\^12\catcode `\_12\catcode `\%12\relax}%
\providecommand \@@startlink[1]{}%
\providecommand \@@endlink[0]{}%
\providecommand \url  [0]{\begingroup\@sanitize@url \@url }%
\providecommand \@url [1]{\endgroup\@href {#1}{\urlprefix }}%
\providecommand \urlprefix  [0]{URL }%
\providecommand \Eprint [0]{\href }%
\providecommand \doibase [0]{https://doi.org/}%
\providecommand \selectlanguage [0]{\@gobble}%
\providecommand \bibinfo  [0]{\@secondoftwo}%
\providecommand \bibfield  [0]{\@secondoftwo}%
\providecommand \translation [1]{[#1]}%
\providecommand \BibitemOpen [0]{}%
\providecommand \bibitemStop [0]{}%
\providecommand \bibitemNoStop [0]{.\EOS\space}%
\providecommand \EOS [0]{\spacefactor3000\relax}%
\providecommand \BibitemShut  [1]{\csname bibitem#1\endcsname}%
\let\auto@bib@innerbib\@empty
%</preamble>
\bibitem [{\citenamefont {Weinberg}(2005)}]{weinberg2005}%
  \BibitemOpen
  \bibfield  {author} {\bibinfo {author} {\bibfnamefont {S.}~\bibnamefont
  {Weinberg}},\ }\href@noop {} {\emph {\bibinfo {title} {The Quantum Theory of
  Fields}}}\ (\bibinfo  {publisher} {Cambridge University Press},\ \bibinfo
  {year} {2005})\BibitemShut {NoStop}%
\bibitem [{\citenamefont {Anderson}(1984)}]{anderson1984}%
  \BibitemOpen
  \bibfield  {author} {\bibinfo {author} {\bibfnamefont {P.~W.}\ \bibnamefont
  {Anderson}},\ }\href@noop {} {\emph {\bibinfo {title} {Basic notations of
  condensed matter physics}}}\ (\bibinfo  {publisher} {The Benjamin/Cummings
  Publishing Company, Inc.},\ \bibinfo {year} {1984})\BibitemShut {NoStop}%
\bibitem [{\citenamefont {Affleck}\ and\ \citenamefont
  {Marston}(1988)}]{affleck1988}%
  \BibitemOpen
  \bibfield  {author} {\bibinfo {author} {\bibfnamefont {I.}~\bibnamefont
  {Affleck}}\ and\ \bibinfo {author} {\bibfnamefont {J.~B.}\ \bibnamefont
  {Marston}},\ }\bibfield  {title} {\bibinfo {title} {Large-n limit of the
  heisenberg-hubbard model: Implications for high-$t_{c}$ superconductors},\
  }\href {https://doi.org/10.1103/PhysRevB.37.3774} {\bibfield  {journal}
  {\bibinfo  {journal} {Phys. Rev. B}\ }\textbf {\bibinfo {volume} {37}},\
  \bibinfo {pages} {3774} (\bibinfo {year} {1988})}\BibitemShut {NoStop}%
\bibitem [{\citenamefont {Harada}\ \emph {et~al.}(2003)\citenamefont {Harada},
  \citenamefont {Kawashima},\ and\ \citenamefont {Troyer}}]{harada2003}%
  \BibitemOpen
  \bibfield  {author} {\bibinfo {author} {\bibfnamefont {K.}~\bibnamefont
  {Harada}}, \bibinfo {author} {\bibfnamefont {N.}~\bibnamefont {Kawashima}},\
  and\ \bibinfo {author} {\bibfnamefont {M.}~\bibnamefont {Troyer}},\
  }\bibfield  {title} {\bibinfo {title} {N{\'e}el and spin-peierls ground
  states of two-dimensional su(n) quantum antiferromagnets},\ }\href
  {https://doi.org/10.1103/PhysRevLett.90.117203} {\bibfield  {journal}
  {\bibinfo  {journal} {Phys. Rev. Lett.}\ }\textbf {\bibinfo {volume} {90}},\
  \bibinfo {pages} {117203} (\bibinfo {year} {2003})}\BibitemShut {NoStop}%
\bibitem [{\citenamefont {Sachdev}\ and\ \citenamefont
  {Read}(1991)}]{sachdev1991}%
  \BibitemOpen
  \bibfield  {author} {\bibinfo {author} {\bibfnamefont {S.}~\bibnamefont
  {Sachdev}}\ and\ \bibinfo {author} {\bibfnamefont {N.}~\bibnamefont {Read}},\
  }\bibfield  {title} {\bibinfo {title} {Large n expansion for frustrated and
  doped quantum antiferromagnets},\ }\href
  {http://www.citebase.org/abstract?id=oai:arXiv.org:cond-mat/0402109}
  {\bibfield  {journal} {\bibinfo  {journal} {Int. J. Mod. Phys. B}\ }\textbf
  {\bibinfo {volume} {5}},\ \bibinfo {pages} {219} (\bibinfo {year}
  {1991})}\BibitemShut {NoStop}%
\bibitem [{\citenamefont {Wu}(2010)}]{wu2010}%
  \BibitemOpen
  \bibfield  {author} {\bibinfo {author} {\bibfnamefont {C.}~\bibnamefont
  {Wu}},\ }\bibfield  {title} {\bibinfo {title} {Exotic many-body physics with
  large-spin fermi gases},\ }\href@noop {} {\bibfield  {journal} {\bibinfo
  {journal} {Physics}\ }\textbf {\bibinfo {volume} {3}},\ \bibinfo {pages} {92}
  (\bibinfo {year} {2010})}\BibitemShut {NoStop}%
\bibitem [{\citenamefont {Wu}(2012)}]{wu2012}%
  \BibitemOpen
  \bibfield  {author} {\bibinfo {author} {\bibfnamefont {C.}~\bibnamefont
  {Wu}},\ }\bibfield  {title} {\bibinfo {title} {Mott made easy},\ }\href
  {https://doi.org/10.1038/nphys2432} {\bibfield  {journal} {\bibinfo
  {journal} {Nature Physics}\ }\textbf {\bibinfo {volume} {8}},\ \bibinfo
  {pages} {784} (\bibinfo {year} {2012})}\BibitemShut {NoStop}%
\bibitem [{\citenamefont {Taie}\ \emph {et~al.}(2012)\citenamefont {Taie},
  \citenamefont {Yamazaki}, \citenamefont {Sugawa},\ and\ \citenamefont
  {Takahashi}}]{taie2012}%
  \BibitemOpen
  \bibfield  {author} {\bibinfo {author} {\bibfnamefont {S.}~\bibnamefont
  {Taie}}, \bibinfo {author} {\bibfnamefont {R.}~\bibnamefont {Yamazaki}},
  \bibinfo {author} {\bibfnamefont {S.}~\bibnamefont {Sugawa}},\ and\ \bibinfo
  {author} {\bibfnamefont {Y.}~\bibnamefont {Takahashi}},\ }\bibfield  {title}
  {\bibinfo {title} {An su(6) mott insulator of an atomic fermi gas realized by
  large-spin pomeranchuk cooling},\ }\href {https://doi.org/10.1038/nphys2430}
  {\bibfield  {journal} {\bibinfo  {journal} {Nature Physics}\ }\textbf
  {\bibinfo {volume} {8}},\ \bibinfo {pages} {825} (\bibinfo {year}
  {2012})}\BibitemShut {NoStop}%
\bibitem [{\citenamefont {DeSalvo}\ \emph {et~al.}(2010)\citenamefont
  {DeSalvo}, \citenamefont {Yan}, \citenamefont {Mickelson}, \citenamefont
  {Martinez~de Escobar},\ and\ \citenamefont {Killian}}]{killian2010}%
  \BibitemOpen
  \bibfield  {author} {\bibinfo {author} {\bibfnamefont {B.~J.}\ \bibnamefont
  {DeSalvo}}, \bibinfo {author} {\bibfnamefont {M.}~\bibnamefont {Yan}},
  \bibinfo {author} {\bibfnamefont {P.~G.}\ \bibnamefont {Mickelson}}, \bibinfo
  {author} {\bibfnamefont {Y.~N.}\ \bibnamefont {Martinez~de Escobar}},\ and\
  \bibinfo {author} {\bibfnamefont {T.~C.}\ \bibnamefont {Killian}},\
  }\bibfield  {title} {\bibinfo {title} {Degenerate fermi gas of
  $^{87}\mathrm{Sr}$},\ }\href {https://doi.org/10.1103/PhysRevLett.105.030402}
  {\bibfield  {journal} {\bibinfo  {journal} {Phys. Rev. Lett.}\ }\textbf
  {\bibinfo {volume} {105}},\ \bibinfo {pages} {030402} (\bibinfo {year}
  {2010})}\BibitemShut {NoStop}%
\bibitem [{\citenamefont {Wu}\ \emph {et~al.}(2003)\citenamefont {Wu},
  \citenamefont {Hu},\ and\ \citenamefont {Zhang}}]{wu2003}%
  \BibitemOpen
  \bibfield  {author} {\bibinfo {author} {\bibfnamefont {C.}~\bibnamefont
  {Wu}}, \bibinfo {author} {\bibfnamefont {J.-p.}\ \bibnamefont {Hu}},\ and\
  \bibinfo {author} {\bibfnamefont {S.-c.}\ \bibnamefont {Zhang}},\ }\bibfield
  {title} {\bibinfo {title} {Exact so(5) symmetry in the spin-$3/2$ fermionic
  system},\ }\href {https://doi.org/10.1103/PhysRevLett.91.186402} {\bibfield
  {journal} {\bibinfo  {journal} {Phys. Rev. Lett.}\ }\textbf {\bibinfo
  {volume} {91}},\ \bibinfo {pages} {186402} (\bibinfo {year}
  {2003})}\BibitemShut {NoStop}%
\bibitem [{\citenamefont {Wu}(2006)}]{wu2006}%
  \BibitemOpen
  \bibfield  {author} {\bibinfo {author} {\bibfnamefont {C.}~\bibnamefont
  {Wu}},\ }\bibfield  {title} {\bibinfo {title} {Hidden symmetry and quantum
  phases in spin-3/2 cold atomic systems},\ }\href
  {https://doi.org/10.1142/s0217984906012213} {\bibfield  {journal} {\bibinfo
  {journal} {Modern Physics Letters B}\ }\textbf {\bibinfo {volume} {20}},\
  \bibinfo {pages} {1707} (\bibinfo {year} {2006})}\BibitemShut {NoStop}%
\bibitem [{\citenamefont {Wu}\ \emph {et~al.}(2010)\citenamefont {Wu},
  \citenamefont {Hu},\ and\ \citenamefont {Zhang}}]{wu2010a}%
  \BibitemOpen
  \bibfield  {author} {\bibinfo {author} {\bibfnamefont {C.}~\bibnamefont
  {Wu}}, \bibinfo {author} {\bibfnamefont {J.}~\bibnamefont {Hu}},\ and\
  \bibinfo {author} {\bibfnamefont {S.-C.}\ \bibnamefont {Zhang}},\ }\bibfield
  {title} {\bibinfo {title} {Quintet pairing and non-abelian vortex string in
  spin-3/2 cold atomic systems},\ }\href
  {https://doi.org/10.1142/s0217979210054968} {\bibfield  {journal} {\bibinfo
  {journal} {International Journal of Modern Physics B}\ }\textbf {\bibinfo
  {volume} {24}},\ \bibinfo {pages} {311} (\bibinfo {year} {2010})}\BibitemShut
  {NoStop}%
\bibitem [{\citenamefont {Gorshkov}\ \emph {et~al.}(2010)\citenamefont
  {Gorshkov}, \citenamefont {Hermele}, \citenamefont {Gurarie}, \citenamefont
  {Xu}, \citenamefont {Julienne}, \citenamefont {Ye}, \citenamefont {Zoller},
  \citenamefont {Demler}, \citenamefont {Lukin},\ and\ \citenamefont
  {Rey}}]{gorshkov2010}%
  \BibitemOpen
  \bibfield  {author} {\bibinfo {author} {\bibfnamefont {A.~V.}\ \bibnamefont
  {Gorshkov}}, \bibinfo {author} {\bibfnamefont {M.}~\bibnamefont {Hermele}},
  \bibinfo {author} {\bibfnamefont {V.}~\bibnamefont {Gurarie}}, \bibinfo
  {author} {\bibfnamefont {C.}~\bibnamefont {Xu}}, \bibinfo {author}
  {\bibfnamefont {P.~S.}\ \bibnamefont {Julienne}}, \bibinfo {author}
  {\bibfnamefont {J.}~\bibnamefont {Ye}}, \bibinfo {author} {\bibfnamefont
  {P.}~\bibnamefont {Zoller}}, \bibinfo {author} {\bibfnamefont
  {E.}~\bibnamefont {Demler}}, \bibinfo {author} {\bibfnamefont {M.~D.}\
  \bibnamefont {Lukin}},\ and\ \bibinfo {author} {\bibfnamefont {A.~M.}\
  \bibnamefont {Rey}},\ }\bibfield  {title} {\bibinfo {title} {Two-orbital
  su(n) magnetism with ultracold alkaline-earth atoms},\ }\href
  {https://doi.org/10.1038/nphys1535} {\bibfield  {journal} {\bibinfo
  {journal} {Nature Phys.}\ }\textbf {\bibinfo {volume} {6}},\ \bibinfo {pages}
  {289} (\bibinfo {year} {2010})}\BibitemShut {NoStop}%
\bibitem [{\citenamefont {Cazalilla}\ \emph {et~al.}(2009)\citenamefont
  {Cazalilla}, \citenamefont {Ho},\ and\ \citenamefont {Ueda}}]{cazalilla2009}%
  \BibitemOpen
  \bibfield  {author} {\bibinfo {author} {\bibfnamefont {M.~A.}\ \bibnamefont
  {Cazalilla}}, \bibinfo {author} {\bibfnamefont {A.~F.}\ \bibnamefont {Ho}},\
  and\ \bibinfo {author} {\bibfnamefont {M.}~\bibnamefont {Ueda}},\ }\bibfield
  {title} {\bibinfo {title} {Ultracold gases of ytterbium: ferromagnetism and
  mott states in an su(6) fermi system},\ }\href
  {https://doi.org/10.1088/1367-2630/11/10/103033} {\bibfield  {journal}
  {\bibinfo  {journal} {New J. Phys.}\ }\textbf {\bibinfo {volume} {11}},\
  \bibinfo {pages} {103033} (\bibinfo {year} {2009})}\BibitemShut {NoStop}%
\bibitem [{\citenamefont {Ozawa}\ \emph {et~al.}(2018)\citenamefont {Ozawa},
  \citenamefont {Taie}, \citenamefont {Takasu},\ and\ \citenamefont
  {Takahashi}}]{ozawa2018}%
  \BibitemOpen
  \bibfield  {author} {\bibinfo {author} {\bibfnamefont {H.}~\bibnamefont
  {Ozawa}}, \bibinfo {author} {\bibfnamefont {S.}~\bibnamefont {Taie}},
  \bibinfo {author} {\bibfnamefont {Y.}~\bibnamefont {Takasu}},\ and\ \bibinfo
  {author} {\bibfnamefont {Y.}~\bibnamefont {Takahashi}},\ }\bibfield  {title}
  {\bibinfo {title} {Antiferromagnetic spin correlation of
  $\mathrm{SU}(\mathcal{N})$ fermi gas in an optical superlattice},\ }\href
  {https://doi.org/10.1103/PhysRevLett.121.225303} {\bibfield  {journal}
  {\bibinfo  {journal} {Phys. Rev. Lett.}\ }\textbf {\bibinfo {volume} {121}},\
  \bibinfo {pages} {225303} (\bibinfo {year} {2018})}\BibitemShut {NoStop}%
\bibitem [{\citenamefont {Taie}\ \emph {et~al.}(2022)\citenamefont {Taie},
  \citenamefont {Ibarra-Garc{\'\i}a-Padilla}, \citenamefont {Nishizawa},
  \citenamefont {Takasu}, \citenamefont {Kuno}, \citenamefont {Wei},
  \citenamefont {Scalettar}, \citenamefont {Hazzard},\ and\ \citenamefont
  {Takahashi}}]{taie2020}%
  \BibitemOpen
  \bibfield  {author} {\bibinfo {author} {\bibfnamefont {S.}~\bibnamefont
  {Taie}}, \bibinfo {author} {\bibfnamefont {E.}~\bibnamefont
  {Ibarra-Garc{\'\i}a-Padilla}}, \bibinfo {author} {\bibfnamefont
  {N.}~\bibnamefont {Nishizawa}}, \bibinfo {author} {\bibfnamefont
  {Y.}~\bibnamefont {Takasu}}, \bibinfo {author} {\bibfnamefont
  {Y.}~\bibnamefont {Kuno}}, \bibinfo {author} {\bibfnamefont {H.-T.}\
  \bibnamefont {Wei}}, \bibinfo {author} {\bibfnamefont {R.~T.}\ \bibnamefont
  {Scalettar}}, \bibinfo {author} {\bibfnamefont {K.~R.~A.}\ \bibnamefont
  {Hazzard}},\ and\ \bibinfo {author} {\bibfnamefont {Y.}~\bibnamefont
  {Takahashi}},\ }\bibfield  {title} {\bibinfo {title} {Observation of
  antiferromagnetic correlations in an ultracold su(n) hubbard model},\ }\href
  {https://doi.org/10.1038/s41567-022-01725-6} {\bibfield  {journal} {\bibinfo
  {journal} {Nature Physics}\ }\textbf {\bibinfo {volume} {18}},\ \bibinfo
  {pages} {1356} (\bibinfo {year} {2022})}\BibitemShut {NoStop}%
\bibitem [{\citenamefont {Zamolodchikov}(1989)}]{zamolodchikov1989}%
  \BibitemOpen
  \bibfield  {author} {\bibinfo {author} {\bibfnamefont {A.~B.}\ \bibnamefont
  {Zamolodchikov}},\ }\bibfield  {title} {\bibinfo {title} {Integrals of motion
  and s-matrix of the (scaled) t = tc ising model with magnetic field},\ }\href
  {https://doi.org/10.1142/s0217751x8900176x} {\bibfield  {journal} {\bibinfo
  {journal} {International Journal of Modern Physics A}\ }\textbf {\bibinfo
  {volume} {04}},\ \bibinfo {pages} {4235} (\bibinfo {year}
  {1989})}\BibitemShut {NoStop}%
\bibitem [{\citenamefont {Coldea}\ \emph {et~al.}(2010)\citenamefont {Coldea},
  \citenamefont {Tennant}, \citenamefont {Wheeler}, \citenamefont {Wawrzynska},
  \citenamefont {Prabhakaran}, \citenamefont {Telling}, \citenamefont
  {Habicht}, \citenamefont {Smeibidl},\ and\ \citenamefont
  {Kiefer}}]{coldea2010}%
  \BibitemOpen
  \bibfield  {author} {\bibinfo {author} {\bibfnamefont {R.}~\bibnamefont
  {Coldea}}, \bibinfo {author} {\bibfnamefont {D.}~\bibnamefont {Tennant}},
  \bibinfo {author} {\bibfnamefont {E.}~\bibnamefont {Wheeler}}, \bibinfo
  {author} {\bibfnamefont {E.}~\bibnamefont {Wawrzynska}}, \bibinfo {author}
  {\bibfnamefont {D.}~\bibnamefont {Prabhakaran}}, \bibinfo {author}
  {\bibfnamefont {M.}~\bibnamefont {Telling}}, \bibinfo {author} {\bibfnamefont
  {K.}~\bibnamefont {Habicht}}, \bibinfo {author} {\bibfnamefont
  {P.}~\bibnamefont {Smeibidl}},\ and\ \bibinfo {author} {\bibfnamefont
  {K.}~\bibnamefont {Kiefer}},\ }\bibfield  {title} {\bibinfo {title} {Quantum
  criticality in an ising chain: Experimental evidence for emergent e8
  symmetry},\ }\href {https://doi.org/10.1126/science.1180085} {\bibfield
  {journal} {\bibinfo  {journal} {Science}\ }\textbf {\bibinfo {volume}
  {327}},\ \bibinfo {pages} {177} (\bibinfo {year} {2010})}\BibitemShut
  {NoStop}%
\bibitem [{\citenamefont {Bernevig}\ \emph {et~al.}(2003)\citenamefont
  {Bernevig}, \citenamefont {Hu}, \citenamefont {Toumbas},\ and\ \citenamefont
  {Zhang}}]{bernevig2003}%
  \BibitemOpen
  \bibfield  {author} {\bibinfo {author} {\bibfnamefont {B.~A.}\ \bibnamefont
  {Bernevig}}, \bibinfo {author} {\bibfnamefont {J.}~\bibnamefont {Hu}},
  \bibinfo {author} {\bibfnamefont {N.}~\bibnamefont {Toumbas}},\ and\ \bibinfo
  {author} {\bibfnamefont {S.-C.}\ \bibnamefont {Zhang}},\ }\bibfield  {title}
  {\bibinfo {title} {Eight-dimensional quantum hall effect and ``octonions''},\
  }\href {https://doi.org/10.1103/PhysRevLett.91.236803} {\bibfield  {journal}
  {\bibinfo  {journal} {Phys. Rev. Lett.}\ }\textbf {\bibinfo {volume} {91}},\
  \bibinfo {pages} {236803} (\bibinfo {year} {2003})}\BibitemShut {NoStop}%
\bibitem [{\citenamefont {Lopes}\ \emph {et~al.}(2019)\citenamefont {Lopes},
  \citenamefont {Quito}, \citenamefont {Han},\ and\ \citenamefont
  {Teo}}]{lopes2019}%
  \BibitemOpen
  \bibfield  {author} {\bibinfo {author} {\bibfnamefont {P.~L.~S.}\
  \bibnamefont {Lopes}}, \bibinfo {author} {\bibfnamefont {V.~L.}\ \bibnamefont
  {Quito}}, \bibinfo {author} {\bibfnamefont {B.}~\bibnamefont {Han}},\ and\
  \bibinfo {author} {\bibfnamefont {J.~C.~Y.}\ \bibnamefont {Teo}},\ }\bibfield
   {title} {\bibinfo {title} {Non-abelian twist to integer quantum hall
  states},\ }\href {https://doi.org/10.1103/PhysRevB.100.085116} {\bibfield
  {journal} {\bibinfo  {journal} {Phys. Rev. B}\ }\textbf {\bibinfo {volume}
  {100}},\ \bibinfo {pages} {085116} (\bibinfo {year} {2019})}\BibitemShut
  {NoStop}%
\bibitem [{\citenamefont {Lim}\ \emph {et~al.}(2023)\citenamefont {Lim},
  \citenamefont {Mulligan},\ and\ \citenamefont {Teo}}]{Lim2023}%
  \BibitemOpen
  \bibfield  {author} {\bibinfo {author} {\bibfnamefont {P.~K.}\ \bibnamefont
  {Lim}}, \bibinfo {author} {\bibfnamefont {M.}~\bibnamefont {Mulligan}},\ and\
  \bibinfo {author} {\bibfnamefont {J.~C.~Y.}\ \bibnamefont {Teo}},\ }\bibfield
   {title} {\bibinfo {title} {Partial fillings of the bosonic ${E}_{8}$ quantum
  hall state},\ }\href {https://doi.org/10.1103/PhysRevB.108.035136} {\bibfield
   {journal} {\bibinfo  {journal} {Phys. Rev. B}\ }\textbf {\bibinfo {volume}
  {108}},\ \bibinfo {pages} {035136} (\bibinfo {year} {2023})}\BibitemShut
  {NoStop}%
\bibitem [{\citenamefont {Agricola}(2008)}]{agricola2008}%
  \BibitemOpen
  \bibfield  {author} {\bibinfo {author} {\bibfnamefont {I.}~\bibnamefont
  {Agricola}},\ }\bibfield  {title} {\bibinfo {title} {Old and new on the
  exceptional group {$G_2$}},\ }\href@noop {} {\bibfield  {journal} {\bibinfo
  {journal} {Notices Amer. Math. Soc.}\ }\textbf {\bibinfo {volume} {55}},\
  \bibinfo {pages} {922} (\bibinfo {year} {2008})}\BibitemShut {NoStop}%
\bibitem [{\citenamefont {Baez}(2002)}]{baez2002}%
  \BibitemOpen
  \bibfield  {author} {\bibinfo {author} {\bibfnamefont {J.~C.}\ \bibnamefont
  {Baez}},\ }\bibfield  {title} {\bibinfo {title} {The octonions},\ }\href
  {https://arxiv.org/abs/math/0105155} {\bibfield  {journal} {\bibinfo
  {journal} {Bull. Am. Math. Soc}\ }\textbf {\bibinfo {volume} {39}},\ \bibinfo
  {pages} {145} (\bibinfo {year} {2002})}\BibitemShut {NoStop}%
\bibitem [{\citenamefont {Holland}\ \emph {et~al.}(2003)\citenamefont
  {Holland}, \citenamefont {Minkowski}, \citenamefont {Pepe},\ and\
  \citenamefont {Wiese}}]{holland2003}%
  \BibitemOpen
  \bibfield  {author} {\bibinfo {author} {\bibfnamefont {K.}~\bibnamefont
  {Holland}}, \bibinfo {author} {\bibfnamefont {P.}~\bibnamefont {Minkowski}},
  \bibinfo {author} {\bibfnamefont {M.}~\bibnamefont {Pepe}},\ and\ \bibinfo
  {author} {\bibfnamefont {U.-J.}\ \bibnamefont {Wiese}},\ }\bibfield  {title}
  {\bibinfo {title} {Exceptional confinement in g(2) gauge theory},\ }\href
  {https://doi.org/10.1016/s0550-3213(03)00571-6} {\bibfield  {journal}
  {\bibinfo  {journal} {Nuclear Physics B}\ }\textbf {\bibinfo {volume}
  {668}},\ \bibinfo {pages} {207} (\bibinfo {year} {2003})}\BibitemShut
  {NoStop}%
\bibitem [{\citenamefont {G{\"u}naydin}\ and\ \citenamefont
  {Ketov}(1996)}]{gunaydin1996}%
  \BibitemOpen
  \bibfield  {author} {\bibinfo {author} {\bibfnamefont {M.}~\bibnamefont
  {G{\"u}naydin}}\ and\ \bibinfo {author} {\bibfnamefont {S.~V.}\ \bibnamefont
  {Ketov}},\ }\bibfield  {title} {\bibinfo {title} {Seven-sphere and the
  exceptional n = 7 and n = 8 superconformal algebras},\ }\href
  {https://doi.org/10.1016/0550-3213(96)00088-0} {\bibfield  {journal}
  {\bibinfo  {journal} {Nuclear Physics B}\ }\textbf {\bibinfo {volume}
  {467}},\ \bibinfo {pages} {215} (\bibinfo {year} {1996})}\BibitemShut
  {NoStop}%
\bibitem [{\citenamefont {Hu}\ and\ \citenamefont {Kane}(2018)}]{Hu2018}%
  \BibitemOpen
  \bibfield  {author} {\bibinfo {author} {\bibfnamefont {Y.}~\bibnamefont
  {Hu}}\ and\ \bibinfo {author} {\bibfnamefont {C.~L.}\ \bibnamefont {Kane}},\
  }\bibfield  {title} {\bibinfo {title} {Fibonacci topological
  superconductor},\ }\href {https://doi.org/10.1103/PhysRevLett.120.066801}
  {\bibfield  {journal} {\bibinfo  {journal} {Phys. Rev. Lett.}\ }\textbf
  {\bibinfo {volume} {120}},\ \bibinfo {pages} {066801} (\bibinfo {year}
  {2018})}\BibitemShut {NoStop}%
\bibitem [{\citenamefont {Zhan}\ \emph {et~al.}(2022)\citenamefont {Zhan},
  \citenamefont {Chen}, \citenamefont {Chen}, \citenamefont {Wang},
  \citenamefont {Yu},\ and\ \citenamefont {Luo}}]{luo2020}%
  \BibitemOpen
  \bibfield  {author} {\bibinfo {author} {\bibfnamefont {Y.-M.}\ \bibnamefont
  {Zhan}}, \bibinfo {author} {\bibfnamefont {Y.-G.}\ \bibnamefont {Chen}},
  \bibinfo {author} {\bibfnamefont {B.}~\bibnamefont {Chen}}, \bibinfo {author}
  {\bibfnamefont {Z.}~\bibnamefont {Wang}}, \bibinfo {author} {\bibfnamefont
  {Y.}~\bibnamefont {Yu}},\ and\ \bibinfo {author} {\bibfnamefont
  {X.}~\bibnamefont {Luo}},\ }\bibfield  {title} {\bibinfo {title} {Universal
  topological quantum computation with strongly correlated majorana edge
  modes},\ }\href {https://doi.org/10.1088/1367-2630/ac5f87} {\bibfield
  {journal} {\bibinfo  {journal} {New Journal of Physics}\ }\textbf {\bibinfo
  {volume} {24}},\ \bibinfo {pages} {043009} (\bibinfo {year}
  {2022})}\BibitemShut {NoStop}%
\bibitem [{\citenamefont {Li}\ \emph {et~al.}(2023)\citenamefont {Li},
  \citenamefont {Quito}, \citenamefont {Schuricht},\ and\ \citenamefont
  {Lopes}}]{Li2023}%
  \BibitemOpen
  \bibfield  {author} {\bibinfo {author} {\bibfnamefont {C.}~\bibnamefont
  {Li}}, \bibinfo {author} {\bibfnamefont {V.~L.}\ \bibnamefont {Quito}},
  \bibinfo {author} {\bibfnamefont {D.}~\bibnamefont {Schuricht}},\ and\
  \bibinfo {author} {\bibfnamefont {P.~L.~S.}\ \bibnamefont {Lopes}},\
  }\bibfield  {title} {\bibinfo {title} {${G}_{2}$ integrable point
  characterization via isotropic spin-3 chains},\ }\href
  {https://doi.org/10.1103/PhysRevB.108.165123} {\bibfield  {journal} {\bibinfo
   {journal} {Phys. Rev. B}\ }\textbf {\bibinfo {volume} {108}},\ \bibinfo
  {pages} {165123} (\bibinfo {year} {2023})}\BibitemShut {NoStop}%
\bibitem [{sup()}]{supp}%
  \BibitemOpen
  \bibinfo {note} {{See Supplemental Material at [URL] for the conventions and
  detailed calculations.}}\BibitemShut {Stop}%
\bibitem [{Note1()}]{Note1}%
  \BibitemOpen
  \bibinfo {note} {An immediately question is what if the rotation in $T$-$V$
  plane is $-120^\circ $, instead of $120^\circ $. In fact, if we define
  $T_a^{\prime \prime }+i V_a^{\prime \prime }= e^{-i\protect \frac {2}{3}\pi }
  (T_a+iV_a)$, and accordingly define $M^{\prime \prime }_{ab}$, it will indeed
  give the third $SO(7)$ symmetry! This third $SO(7)$ symmetry is self-dual
  under $\chi _0(\protect \mathbf {i})$, namely $\chi _0(\protect \mathbf {i})
  M^{\prime \prime }_{ab}(\protect \mathbf {i})\chi _0(\protect \mathbf
  {i})=M^{\prime \prime }_{ab}(\protect \mathbf {i})$ and $\chi _0(\protect
  \mathbf {i}) V^{\prime \prime }_{a}(\protect \mathbf {i})\chi _0(\protect
  \mathbf {i})=-V^{\prime \prime }_{a}(\protect \mathbf {i})$. However, a
  straightforward algebra shows its Casimir equal to a $c$-number, 21, instead
  of some operator in terms of $\psi _\sigma (\protect \mathbf {i})$.
  Therefore, this third $SO(7)$ symmetry is absent in the lattice model.
  Nevertheless, in a sequential work~\cite {Gao2024} we show that it can emerge
  in the continuum limit, and fulfill the triality structure of $SO(8)$
  together with $SO(7)_A$ and $SO(7)_B$.}\BibitemShut {Stop}%
\bibitem [{\citenamefont {Gao}\ and\ \citenamefont {Wu}(2024)}]{Gao2024}%
  \BibitemOpen
  \bibfield  {author} {\bibinfo {author} {\bibfnamefont {Z.-Q.}\ \bibnamefont
  {Gao}}\ and\ \bibinfo {author} {\bibfnamefont {C.}~\bibnamefont {Wu}},\
  }\href {https://arxiv.org/abs/2411.08107} {\bibinfo {title} {From $g_2$ to
  $so(8)$: Emergence and reminiscence of supersymmetry and triality}} (\bibinfo
  {year} {2024}),\ \Eprint {https://arxiv.org/abs/2411.08107} {arXiv:2411.08107
  [cond-mat.str-el]} \BibitemShut {NoStop}%
\bibitem [{\citenamefont {Agricola}\ \emph {et~al.}(2018)\citenamefont
  {Agricola}, \citenamefont {Bor{\'o}wka},\ and\ \citenamefont
  {Friedrich}}]{agricola2018}%
  \BibitemOpen
  \bibfield  {author} {\bibinfo {author} {\bibfnamefont {I.}~\bibnamefont
  {Agricola}}, \bibinfo {author} {\bibfnamefont {A.}~\bibnamefont
  {Bor{\'o}wka}},\ and\ \bibinfo {author} {\bibfnamefont {T.}~\bibnamefont
  {Friedrich}},\ }\bibfield  {title} {\bibinfo {title} {S6 and the geometry of
  nearly k{\"a}hler 6-manifolds},\ }\href
  {https://doi.org/https://doi.org/10.1016/j.difgeo.2017.10.007} {\bibfield
  {journal} {\bibinfo  {journal} {Differential Geometry and its Applications}\
  }\textbf {\bibinfo {volume} {57}},\ \bibinfo {pages} {75 } (\bibinfo {year}
  {2018})},\ \bibinfo {note} {(Non)-existence of complex structures on
  S6}\BibitemShut {NoStop}%
\bibitem [{Note2()}]{Note2}%
  \BibitemOpen
  \bibinfo {note} {In the sequential work~\cite {Gao2024} we show that in
  (1+1)D spacetime supersymmetry can emerge as a tricritical Ising conformal
  field theory with central charge $c=\protect \frac {7}{10}$. In fact, our
  model with $u=v$ reproduces the (lattice version of) Fibonacci topological
  field theory constructed by 7 interacting Majorana fermions~\cite {Hu2018},
  while the latter one is shown to indeed have central charge $c=\protect \frac
  {7}{10}$.}\BibitemShut {Stop}%
\bibitem [{\citenamefont {Grover}\ \emph {et~al.}(2014)\citenamefont {Grover},
  \citenamefont {Sheng},\ and\ \citenamefont {Vishwanath}}]{Grover2014}%
  \BibitemOpen
  \bibfield  {author} {\bibinfo {author} {\bibfnamefont {T.}~\bibnamefont
  {Grover}}, \bibinfo {author} {\bibfnamefont {D.~N.}\ \bibnamefont {Sheng}},\
  and\ \bibinfo {author} {\bibfnamefont {A.}~\bibnamefont {Vishwanath}},\
  }\bibfield  {title} {\bibinfo {title} {Emergent space-time supersymmetry at
  the boundary of a topological phase},\ }\href
  {https://doi.org/10.1126/science.1248253} {\bibfield  {journal} {\bibinfo
  {journal} {Science}\ }\textbf {\bibinfo {volume} {344}},\ \bibinfo {pages}
  {280} (\bibinfo {year} {2014})}\BibitemShut {NoStop}%
\bibitem [{Note3()}]{Note3}%
  \BibitemOpen
  \bibinfo {note} {For $G_2$, another kind of important symmetry breaking
  pattern is $G_2$ broken to $SU(2)\times SU(2)$. However, in our model, this
  symmetry breaking pattern does not appear naturally.}\BibitemShut {Stop}%
\end{thebibliography}%
\onecolumngrid

\renewcommand\theequation{S\arabic{equation}}
\renewcommand\thefigure{S\arabic{figure}}
\renewcommand\bibnumfmt[1]{[S#1]}
\setcounter{equation}{0}
\setcounter{figure}{0}

\clearpage{}
\begin{center}
\Large{\bf Supplemental Material for ``Construction of $G_2$ symmetry in a Hubbard-type model"}
\end{center}

\section{$G_2$ theory is interacting}
It is worth noticing that $G_2$ symmetry is intrinsically strongly correlated: an internal $G_2$ symmetry (acting in internal degrees of freedom and unrelated to spacetime) cannot be realized in non-interacting systems. To see this, suppose a general non-interacting Hamiltonian $H=\sum_{\mathbf{x},\mathbf{x}^\prime}\Phi^\dag(\mathbf{x})\mathcal{H}(\mathbf{x},\mathbf{x}^\prime)\Phi(\mathbf{x}^\prime)$ has an internal $G_2$ symmetry, where $\Phi(\mathbf{x})$ defined at space coordinate $\mathbf{x}$ can be a multi-component boson, complex fermion, or Majorana fermion field. The spacetime independence of the $G_2$ symmetry ensures that the $G_2$ operations are not involved with $\mathbf{x}$. Without losing of generality, we assume that $\Phi(\mathbf{x})$ transforms under non-trivial $G_2$ irreducible representation $R_\Phi$ with dimension $d_\Phi$, and hence $d_\Phi\ge 7$. Therefore, according to Schur's lemma, $\mathcal{H}(\mathbf{x},\mathbf{x}^\prime)$ must be proportional to the $d_\Phi$-dimensional identity matrix, $\mathcal{H}(\mathbf{x},\mathbf{x}^\prime)=\mathbb{I}_{d_\Phi}f(\mathbf{x},\mathbf{x}^\prime)$, where $f(\mathbf{x},\mathbf{x}^\prime)$ is a scalar function. However, the symmetry of Hamiltonian $H=\sum_{\mathbf{x},\mathbf{x}^\prime}f(\mathbf{x},\mathbf{x}^\prime)\Phi^\dag(\mathbf{x})\Phi(\mathbf{x}^\prime)$ is not $G_2$, but enlarged to $SU(d_\Phi)$ or $SO(d_\Phi)$ for complex or real field $\Phi(\mathbf{x})$, respectively, either of which contains $G_2$ (actually $SO(7)$ since $d_\Phi\ge 7$) as a subgroup. Therefore, whenever a non-interacting system has an internal $G_2$ symmetry, it will be at least $SO(7)$ symmetric. To construct a maximally $G_2$ symmetric model, one inevitably needs to introduce interactions. This argument can be generalized that any Lie group which is a subgroup of $SO(d_0)$, where $d_0$ is the dimension of its smallest non-trivial irreducible representation, is intrinsically strongly correlated. Therefore, exceptional Lie groups are all intrinsically strongly correlated, but classical Lie groups are not.

\section{$SO(8)$ algebra in 4-component fermion systems}

In this section we briefly review the $SO(8)$ algebra in 4-component spin-$\frac{3}{2}$ fermion systems. We adopt the convention for the four by four gamma matrices as
\beq
\Gamma^1 =\begin{pmatrix}
0 & -iI\\
iI & 0
\end{pmatrix},\quad
\Gamma^{2,3,4} =\begin{pmatrix}
\vec{\sigma} & 0\\
0 & -\vec{\sigma}
\end{pmatrix},\quad
\Gamma^5 =\begin{pmatrix}
0 & I\\
I & 0
\end{pmatrix},
\eeq
where $I$ and $\vec{\sigma}$ are two by two identity and Pauli matrices. These five gamma matrices form an $Sp(4)$ vector. The generators of $Sp(4)$ are defined as $\Gamma^{ab}=-\frac{i}{2}[\Gamma^a, \Gamma^b]$, and the charge conjugation operator is defined as $R=\Gamma^1\Gamma^3$. Therefore, the fermion bilinear operators $M_{ab}(\mathbf{i})$ with $0\le a, b\le 7$:
\beq
&&M_{ab}(\mathbf{i})=L_{ab}(\mathbf{i})=-\frac{1}{2}\psi^\dagger(\mathbf{i}) \Gamma^{ab} \psi(\mathbf{i}),\quad 1\le a, b\le 5,\\
&&M_{a7}(\mathbf{i})=n_a(\mathbf{i})=\frac{1}{2}\psi^\dagger(\mathbf{i}) \Gamma^a \psi(\mathbf{i}),\quad 1\le a\le 5,\\
&&M_{06}(\mathbf{i})=N(\mathbf{i})=\frac{1}{2}\psi^\dagger(\mathbf{i}) \psi(\mathbf{i})-1,\\
&&M_{0a}(\mathbf{i})+iM_{a6}(\mathbf{i})=\text{Re}\xi_a(\mathbf{i})+i\text{Im}\xi_a(\mathbf{i})=-\frac{i}{2}\text{sgn}(\mathbf{i})\psi^\dagger (\mathbf{i}) \Gamma^a R \psi^\dagger(\mathbf{i}),\quad 1\le a\le 5,\\
&&M_{07}(\mathbf{i})-iM_{67}(\mathbf{i})=\text{Re}\eta(\mathbf{i})+i\text{Im}\eta(\mathbf{i})=\frac{1}{2}\text{sgn}(\mathbf{i})\psi^\dagger (\mathbf{i}) R \psi^\dagger(\mathbf{i}),
\eeq
span an $SO(8)$ algebra, which can be checked straightforwardly by the commutation rule
\beq
[M_{ab}(\mathbf{i}),M_{cd}(\mathbf{i})]=i(\delta_{ad}M_{bc}(\mathbf{i})+\delta_{bc}M_{ad}(\mathbf{i})-\delta_{ac}M_{bd}(\mathbf{i})-\delta_{bd}M_{ac}(\mathbf{i})), \quad 0\le a, b, c, d\le 7.
\eeq
The $SO(8)$ generators can be formally written as
\beq
M_{ab}(\mathbf{i})=\begin{pmatrix}
0 & \text{Re}\xi_1(\mathbf{i}) \sim \text{Re}\xi_5(\mathbf{i}) & N(\mathbf{i}) & \text{Re}\eta(\mathbf{i}) \\
& & \text{Im}\xi_1(\mathbf{i}) & n_1(\mathbf{i})\\
& L_{ab}(\mathbf{i}) & \sim & \sim \\
& & \text{Im}\xi_5(\mathbf{i}) & n_5(\mathbf{i})\\
& & 0 & -\text{Im}\eta(\mathbf{i})\\
& & & 0
\end{pmatrix},
\eeq
where $L_{ab}(\mathbf{i})$ span an $Sp(4)$ algebra, and $L_{ab}(\mathbf{i})$ together with $\xi_a(\mathbf{i})$ and $N(\mathbf{i})$ span the $SO(7)_A$ algebra discussed in the main text. Matrix form of $SO(7)_A$ generators $M_{ab}$ with $0\le a,b\le 6$ (neglected normalizations) in Nambu basis reads
\begin{subequations}
\begin{align}
&M_{ab}=\begin{pmatrix}
\Gamma_{ab} & 0\\
0 & -\Gamma_{ab}^T
\end{pmatrix},~
1\le a,b\le 5,\\
&M_{0a}=\begin{pmatrix}
0 & -i\Gamma_a R\\
-i\Gamma_a ^T R &0
\end{pmatrix},~
1\le a\le 5,\\
&M_{a6}=\begin{pmatrix}
0 & -\Gamma_a R\\
\Gamma_a ^T R &0
\end{pmatrix},~
1\le a\le 5,\\
&M_{06}=\begin{pmatrix}
I & 0\\
0 & -I
\end{pmatrix}.
\end{align}
\end{subequations}
By means of $T_a(\mathbf{i})=\frac{1}{2\sqrt{3}} C_{abc}  M_{bc}(\mathbf{i})$ and $C_{031} = C_{052} = C_{064} = C_{126} = C_{154} = C_{243} = C_{365} = 1$, the matrix form of $T_a$ is
\begin{subequations}
\begin{align}
&T_0=\frac{1}{\sqrt{3}}\begin{pmatrix}
-\Gamma_{13}-\Gamma_{25} & \Gamma_4 R\\
-\Gamma_4 ^T R & \Gamma_{13}^T+\Gamma_{25}^T
\end{pmatrix},\\
&T_1=\frac{1}{\sqrt{3}}\begin{pmatrix}
-\Gamma_{45} & -\Gamma_2 R-i\Gamma_3 R\\
\Gamma_2 ^T R-i\Gamma_3 ^T R & \Gamma_{45}^T
\end{pmatrix},\\
&T_2=\frac{1}{\sqrt{3}}\begin{pmatrix}
-\Gamma_{34} & \Gamma_1 R-i\Gamma_5 R\\
-\Gamma_1 ^T R-i\Gamma_5 ^T R & \Gamma_{34}^T
\end{pmatrix},\\
&T_3=\frac{1}{\sqrt{3}}\begin{pmatrix}
\Gamma_{24} & i\Gamma_1 R+\Gamma_5 R\\
i\Gamma_1 ^T R-\Gamma_5 ^T R & -\Gamma_{24}^T
\end{pmatrix},\\
&T_4=\frac{1}{\sqrt{3}}\begin{pmatrix}
\Gamma_{15}-\Gamma_{23}+I & 0\\
0 & -\Gamma_{15}^T+\Gamma_{23}^T-I
\end{pmatrix},~~~\mathrm{(diagonal)},\\
&T_5=\frac{1}{\sqrt{3}}\begin{pmatrix}
-\Gamma_{14} & -\Gamma_3 R+i\Gamma_2 R\\
\Gamma_3 ^T R+i\Gamma_2 ^T R & \Gamma_{14}^T
\end{pmatrix},\\
&T_6=\frac{1}{\sqrt{3}}\begin{pmatrix}
\Gamma_{12}-\Gamma_{35} & i\Gamma_4 R\\
i\Gamma_4^T R & -\Gamma_{12}^T+\Gamma_{35}^T
\end{pmatrix}.
\end{align}
\end{subequations}

\section{$G_2$ algebra}

\begin{figure*}[htbp]
\centering
\includegraphics[width=0.3\linewidth]{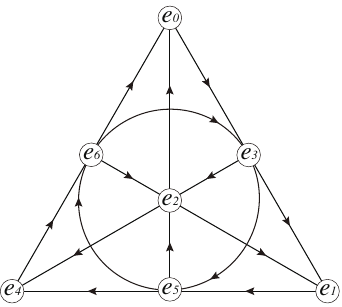}
\caption{A Fano plane compatible with our convention of $C_{abc}$.}\label{fig:fano}
\end{figure*}

The structure constants of octonion have relations
\begin{subequations}
\begin{align}
&C_{abc}C_{abd}=6\delta_{cd},\\
&C_{abc}C_{abde}=-4C_{cde},\\
&C_{abcd}C_{abef}=-2C_{cdef}+4(\delta_{ce}\delta_{df}-\delta_{cf}\delta_{de}),\\
&C_{abcd}C_{abce}=24\delta_{de}.
\end{align}
\end{subequations}
A set of legible $C_{abc}$ should be consistent with a Fano plane. Along each arrowed line or circle in a Fano plane $C_{abc}=1$, otherwise $C_{abc}=0$, and $C_{abc}$ are fully anti-symmetric. For example, in Fig.~[\ref{fig:fano}], all the non-vanishing positive-valued $C_{abc}$ are $C_{031} = C_{052} = C_{046} = C_{162} = C_{154} = C_{243} = C_{356} = 1$. Other sets of $C_{abc}$ can be generated from Fig.~[\ref{fig:fano}]. All the allowed operations to the Fano plane are permutations of octonion imaginary units $\hat{e}_a$, and reverse the orientations of three concurrent lines and/or circle which means set one $\hat{e}_a$ to $-\hat{e}_a$. For Fig.~[\ref{fig:fano}], if arrows on lines and circle concurrent at $\hat{e}_6$ are reversed, we arrive at the convention of $C_{abc}$ in the main text.

The $G_2$ generators $G_{ab}(\mathbf{i})$ and two sets of $G_2$ vectors $T_a(\mathbf{i})$, $V_a(\mathbf{i})$ have commutators 
\beq
&&[G_{ab}(\mathbf{i}),G_{cd}(\mathbf{i})]=\frac{i}{3}(4\delta_{c[b}G_{a]d}(\mathbf{i})-4\delta_{d[b}G_{a]c}(\mathbf{i})+C_{cde[a}G_{be]}(\mathbf{i})-C_{abe[c}G_{de]}(\mathbf{i})),\\
&&\left[G_{ab}(\mathbf{i}),T_c(\mathbf{i})\right]=\frac{i}{3}(2\delta_{bc}T_a(\mathbf{i})-2\delta_{ac}T_b(\mathbf{i})-C_{abcd}T_d(\mathbf{i})),\\
&&\left[G_{ab}(\mathbf{i}),V_c(\mathbf{i})\right]=\frac{i}{3}(2\delta_{bc}V_a(\mathbf{i})-2\delta_{ac}V_b(\mathbf{i})-C_{abcd}V_d(\mathbf{i})),\\
&&\left[T_a(\mathbf{i}),T_b(\mathbf{i})\right]=i\Big(-G_{ab}(\mathbf{i})+\frac{1}{\sqrt{3}}C_{abc}T_c(\mathbf{i})\Big),\\
&&\left[T_a(\mathbf{i}),V_b(\mathbf{i})\right]=-\frac{i}{\sqrt{3}}C_{abc}V_c(\mathbf{i}),\\
&&\left[V_a(\mathbf{i}),V_b(\mathbf{i})\right]=-iM_{ab}(\mathbf{i}),
\eeq
where subscript $_{[...]}$ stands for antisymmetrization $A_{[ab]}=\frac{1}{2}(A_{ab}-A_{ba})$. These relations also hold if $T_a(\mathbf{i})$ and $V_a(\mathbf{i})$ are substituted to $T^\prime_a(\mathbf{i})$ and $V^\prime_a(\mathbf{i})$ corresponding to $SO(7)_B$, where $T_a^\prime+i V_a^\prime= e^{i\frac{2}{3}\pi} (T_a+iV_a)$. Here we give an explicit proof:
\beq
\left[T_a^\prime ,T_b^\prime\right]&=&\frac{1}{4}\left[T_a,T_b\right]+\frac{3}{4}\left[V_a,V_b\right]+\frac{\sqrt{3}}{4}(\left[T_a,V_b\right]+\left[V_a,T_b\right]\nn\\
&=&i\Big(-\frac{1}{4}G_{ab}+\frac{1}{4\sqrt{3}}C_{abc}T_c-\frac{3}{4}M_{ab}-\frac{1}{2}C_{abc}V_c\Big)\nn\\
&=&i\Big(-\frac{1}{4}G_{ab}+\frac{1}{4\sqrt{3}}C_{abc}T_c-\frac{3}{4}G_{ab}-\frac{3}{4\sqrt{3}}C_{abc}T_c-\frac{1}{2}C_{abc}V_c\Big)\nn\\
&=&i\left(-G_{ab}+\frac{1}{\sqrt{3}}C_{abc}\Big(-\frac{1}{2}T_c-\frac{\sqrt{3}}{2}V_c\Big)\right)\nn\\
&=&i\Big(-G_{ab}+\frac{1}{\sqrt{3}}C_{abc}T_c^\prime\Big),
\eeq
\beq
\left[T_a^\prime ,V_b^\prime\right]&=&-\frac{\sqrt{3}}{4}\left[T_a,T_b\right]+\frac{\sqrt{3}}{4}\left[V_a,V_b\right]+\frac{1}{4}\left[T_a,V_b\right]-\frac{3}{4}\left[V_a,T_b\right]\nn\\
&=&i\Big(\frac{\sqrt{3}}{4}G_{ab}-\frac{1}{4}C_{abc}T_c-\frac{\sqrt{3}}{4}M_{ab}+\frac{1}{2\sqrt{3}}C_{abc}V_c\Big)\nn\\
&=&i\Big(\frac{\sqrt{3}}{4}G_{ab}-\frac{1}{4}C_{abc}T_c-\frac{\sqrt{3}}{4}G_{ab}-\frac{1}{4}C_{abc}T_c+\frac{1}{2\sqrt{3}}C_{abc}V_c\Big)\nn\\
&=&-\frac{i}{\sqrt{3}}C_{abc}\Big(\frac{\sqrt{3}}{2}T_c-\frac{1}{2}V_c\Big)\nn\\
&=&-\frac{i}{\sqrt{3}}C_{abc}V_c^\prime,
\eeq
\beq
\left[V_a^\prime ,V_b^\prime\right]&=&\frac{3}{4}\left[T_a,T_b\right]+\frac{1}{4}\left[V_a,V_b\right]-\frac{\sqrt{3}}{4}(\left[T_a,V_b\right]+\left[V_a,T_b\right])\nn\\
&=&i\Big(-\frac{3}{4}G_{ab}+\frac{\sqrt{3}}{4}C_{abc}T_c-\frac{1}{4}M_{ab}+\frac{1}{2}C_{abc}V_c\Big)\nn\\
&=&i\Big(-\frac{3}{4}G_{ab}+\frac{\sqrt{3}}{4}C_{abc}T_c-\frac{1}{4}G_{ab}-\frac{1}{4\sqrt{3}}C_{abc}T_c+\frac{1}{2}C_{abc}V_c\Big)\nn\\
&=&-i\left(G_{ab}+\frac{1}{\sqrt{3}}C_{abc}\Big(-\frac{1}{2}T_c-\frac{\sqrt{3}}{2}V_c\Big)\right)\nn\\
&=&-i\Big(G_{ab}+\frac{1}{\sqrt{3}}C_{abc}T_c^\prime\Big)\nn\\
&=&-iM_{ab}^\prime,
\eeq
where site index $\mathbf{i}$ is neglected for clarity. Commutators between $SO(7)_B$ generators $M_{ab}^\prime$ and vector $V_a^\prime$ are of the same form of commutators between $M_{ab}$ and $V_a$, which can be also checked straightforwardly.

\begin{figure*}[htbp]
\centering
\includegraphics[width=15cm]{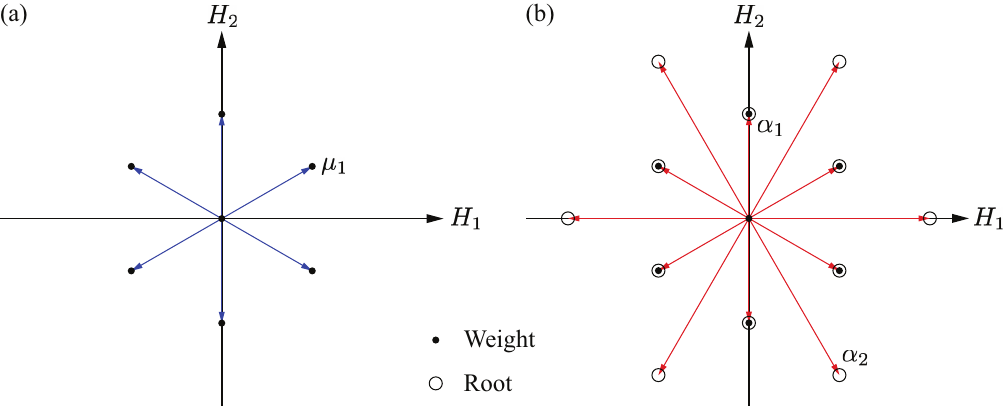}
\caption{(a) The weight diagram. (b) The root diagram.}\label{fig2}
\end{figure*}

The Cartan subalgebra of $G_2$ are chosen as
\beq
H_1=\frac{\sqrt 3}{2}\sum_\mathbf{i} G_{15}(\mathbf{i})+G_{23}(\mathbf{i}), \quad H_2=\frac{3}{2}\sum_\mathbf{i} G_{06}(\mathbf{i}),
\eeq
where $G_{15}$, $G_{23}$ and $G_{06}$ are all diagonal in the fermion basis:
\beq
G_{15}(\mathbf{i})&=&\frac{1}{3}\big(1+n_{\frac{1}{2}}(\mathbf{i})-2n_{-\frac{1}{2}}(\mathbf{i})-n_{-\frac{3}{2}}(\mathbf{i})\big),\quad G_{23}(\mathbf{i})=\frac{1}{3}\big(-1+2n_{\frac{1}{2}}(\mathbf{i})-n_{-\frac{1}{2}}(\mathbf{i})+n_{-\frac{3}{2}}(\mathbf{i})\big),\nn\\
G_{06}(\mathbf{i})&=&\frac{1}{3}\big(-2+n_{\frac{1}{2}}(\mathbf{i})+n_{-\frac{1}{2}}(\mathbf{i})+2n_{-\frac{3}{2}}(\mathbf{i})\big).
\eeq
The conserved quantities can be chosen with explicit physical meanings
\beq
Q_1&=&\frac{2}{\sqrt{3}}H_1=\sum_\mathbf{i} n_{\frac{1}{2}}(\mathbf{i})-n_{-\frac{1}{2}}(\mathbf{i}),\quad Q_2=-\frac{1}{\sqrt{3}}H_1+H_2=\sum_\mathbf{i} n_{-\frac{3}{2}}(\mathbf{i})+n_{-\frac{1}{2}}(\mathbf{i}).
\eeq
Besides, $T_4$ and $V_4$ are also diagonal:
\beq
T_4(\mathbf{i})=\frac{1}{2\sqrt{3}}\big(-2+3n_{\frac{3}{2}}(\mathbf{i})+n_{\frac{1}{2}}(\mathbf{i})+n_{-\frac{1}{2}}(\mathbf{i})-n_{-\frac{3}{2}}(\mathbf{i})\big),\quad V_4(\mathbf{i})=\frac{1}{2}\big(n_{\frac{3}{2}}(\mathbf{i})-n_{\frac{1}{2}}(\mathbf{i})-n_{-\frac{1}{2}}(\mathbf{i})+n_{-\frac{3}{2}}(\mathbf{i})\big).
\eeq
Weight diagram of the 7-dimensional vector representation is in the standard form with the highest weight $\mu_1=(\frac{\sqrt{3}}{2},\frac{1}{2})$, as shown in Fig.~[\ref{fig2} (a)].

Simple roots are chosen as standard form $\alpha_1=(0,1)$, $\alpha_2=(\frac{\sqrt{3}}{2},-\frac{3}{2})$. The root diagram is shown in Fig.~[\ref{fig2} (b)]. The generators in Cartan-Weyl basis and their corresponding roots are listed in Table~\ref{table4}.
\begin{table}[!htbp]
\begin{tabular}{|c|c|c|}
\hline
Root & Cartan-Weyl Basis \\
\hline
$\alpha_1=(0,1)$ & $\frac{\sqrt{6}}{2}(G_{12}-G_{35}+iG_{46})$\\
\hline
$(\frac{-\sqrt{3}}{2},\frac{1}{2})$ & $\frac{\sqrt{6}}{2}(G_{14}-iG_{45})$\\
\hline
$(\frac{-\sqrt{3}}{2},-\frac{1}{2})$ & $\frac{\sqrt{6}}{2}(G_{24}+iG_{34})$\\
\hline
$(0,-1)$ & $\frac{\sqrt{6}}{2}(G_{12}-G_{35}-iG_{46})$\\
\hline
$(\frac{\sqrt{3}}{2},-\frac{1}{2})$ & $\frac{\sqrt{6}}{2}(G_{14}+iG_{45})$\\
\hline
$(\frac{\sqrt{3}}{2},\frac{1}{2})$ & $\frac{\sqrt{6}}{2}(G_{24}-iG_{34})$\\
\hline
$\alpha_2=(\frac{\sqrt{3}}{2},-\frac{3}{2})$ & $\frac{1}{\sqrt{2}}(2G_{16}+iG_{24}+G_{34}-2iG_{56})$\\
\hline
$(\sqrt{3},0)$ & $\frac{1}{\sqrt{2}}(G_{12}+2iG_{25}+G_{35}+iG_{46})$\\
\hline
$(\frac{\sqrt{3}}{2},\frac{3}{2})$ & $\frac{1}{\sqrt{2}}(G_{14}-2iG_{26}-2G_{36}+iG_{45})$\\
\hline
$(-\frac{\sqrt{3}}{2},\frac{3}{2})$ & $\frac{1}{\sqrt{2}}(2G_{16}-iG_{24}+G_{34}+2iG_{56})$\\
\hline
$(-\sqrt{3},0)$ & $\frac{1}{\sqrt{2}}(G_{12}-2iG_{25}+G_{35}-iG_{46})$\\
\hline
$(-\frac{\sqrt{3}}{2},-\frac{3}{2})$ & $\frac{1}{\sqrt{2}}(G_{14}+2iG_{26}-2G_{36}-iG_{45})$\\
\hline
\end{tabular}
\caption{Cartan-Weyl basis and corresponding roots.}\label{table4}
\centering
\end{table}

\section{Explicit form of the interaction Hamiltonian}

The explicit form of the $v$-term of interaction Hamiltonian $H^{G_2}_\text{int}$ reads
\beq
H^{G_2}_{\text{int},v}=2v\sum_\mathbf{i} &-&\text{Im}\xi_1(\mathbf{i})L_{34}(\mathbf{i})+\text{Im}\xi_3(\mathbf{i})L_{14}(\mathbf{i})-\text{Im}\xi_4(\mathbf{i})L_{13}(\mathbf{i})+\text{Im}\xi_2(\mathbf{i})L_{45}(\mathbf{i})-\text{Im}\xi_4(\mathbf{i})L_{25}(\mathbf{i})+\text{Im}\xi_5(\mathbf{i})L_{24}(\mathbf{i})\nn\\
&+&\text{Re}\xi_4(\mathbf{i})L_{12}(\mathbf{i})-\text{Re}\xi_2(\mathbf{i})L_{14}(\mathbf{i})+\text{Re}\xi_1(\mathbf{i})L_{24}(\mathbf{i})+\text{Re}\xi_5(\mathbf{i})L_{34}(\mathbf{i})-\text{Re}\xi_4(\mathbf{i})L_{35}(\mathbf{i})+\text{Re}\xi_3(\mathbf{i})L_{45}(\mathbf{i})\nn\\
&+&\text{Re}\xi_5(\mathbf{i})\text{Im}\xi_1(\mathbf{i})-\text{Re}\xi_1(\mathbf{i})\text{Im}\xi_5(\mathbf{i})+\text{Re}\xi_2(\mathbf{i})\text{Im}\xi_3(\mathbf{i})-\text{Re}\xi_3(\mathbf{i})\text{Im}\xi_2(\mathbf{i})\nn\\
&-&L_{15}(\mathbf{i})N(\mathbf{i})+L_{23}(\mathbf{i})N(\mathbf{i})+L_{15}(\mathbf{i})L_{23}(\mathbf{i})-L_{13}(\mathbf{i})L_{25}(\mathbf{i})+L_{12}(\mathbf{i})L_{35}(\mathbf{i}).
\eeq
Actually, this horrible interaction is much neater in terms of fermion operators. The total interaction Hamiltonian reads
\beq
H^{G_2}_{\text{int}}=\sum_\mathbf{i} &-&\frac{3}{2}(u-v)(n_{\frac{3}{2}}(\mathbf{i})+n_{\frac{1}{2}}(\mathbf{i})+n_{-\frac{1}{2}}(\mathbf{i}))-\frac{3}{2}(u+3v)n_{-\frac{3}{2}}(\mathbf{i})\nn\\
&+&3(u-v)\big(n_{\frac{3}{2}}(\mathbf{i})n_{\frac{1}{2}}(\mathbf{i})+n_{\frac{3}{2}}(\mathbf{i})n_{-\frac{1}{2}}(\mathbf{i})-n_{\frac{3}{2}}(\mathbf{i})n_{-\frac{3}{2}}(\mathbf{i})\big)\nn\\
&+&3(u+v)\big(-n_{\frac{1}{2}}(\mathbf{i})n_{-\frac{1}{2}}(\mathbf{i})+n_{\frac{1}{2}}(\mathbf{i})n_{-\frac{3}{2}}(\mathbf{i})+n_{-\frac{1}{2}}(\mathbf{i})n_{-\frac{3}{2}}(\mathbf{i})\big)\nn\\
&+&6u\big(\psi_{\frac{1}{2}}^\dagger(\mathbf{i})\psi_{-\frac{1}{2}}^\dagger(\mathbf{i})\psi_{-\frac{3}{2}}(\mathbf{i})\psi_{\frac{3}{2}}(\mathbf{i})+h.c.\big)+6iv\text{ sgn}(\mathbf{i})\big(\psi_{\frac{3}{2}}^\dagger(\mathbf{i})\psi_{\frac{1}{2}}^\dagger(\mathbf{i})\psi_{-\frac{1}{2}}^\dagger(\mathbf{i})\psi_{-\frac{3}{2}}(\mathbf{i})-h.c.\big),\label{eq:32}
\eeq
where it can be seen that only the last $6iv$-term violates the conservation of the total particle number.

\section{Mean-Field Decompositions}

In this section we derive the mean-field Hamiltonian explicitly. Certain operator identities are utilized in the calculation of mean-field decomposition.
\begin{subequations}
\begin{align}
C_{abcd}M_{ab}M_{cd}=2G_{ab}G_{ab}-8T_aT_a, &~~~ 
C_{abcd}M'_{ab}M'_{cd}=2G_{ab}G_{ab}-8T'_aT'_a,\label{eq:51}\\
M_{ab}M_{ab}=G_{ab}G_{ab}+2T_aT_a , &~~~   
M'_{ab}M'_{ab}=G_{ab}G_{ab}+2T'_aT'_a,\label{eq:52}\\
V'_aV'_a=2V_aV_a-3T'_aT'_a , &~~~   
V_aV_a=2V'_aV'_a-3T_aT_a,\label{eq:53}\\
14=2V_aV_a+M_{ab}M_{ab}, &~~~   
14=2V'_aV'_a+M'_{ab}M'_{ab}\label{eq:54}.
\end{align}
\end{subequations}
Each identity above should be understood up to a constant. In each quadrant of the single site phase diagram Fig. [1] in main text, the Hamiltonian $H_{G_2}$ can be rearranged as
\beq
H_{G_2,\text{I}}&=&\sum_\mathbf{i} -2|u|V_a(\mathbf{i})V_a(\mathbf{i})-2|v|V'_a(\mathbf{i})V'_a(\mathbf{i})\nn\\
&=&-\sum_\mathbf{i} \frac{2}{|u|+|v|}\big(|u|V_a(\mathbf{i})-|v|V'_a(\mathbf{i})\big)^2,\\
H_{G_2,\text{II}}&=&\sum_\mathbf{i} -|u|M_{ab}(\mathbf{i})M_{ab}(\mathbf{i})-2|v|V'_a(\mathbf{i})V'_a(\mathbf{i})\nn\\
&=&\sum_\mathbf{i} -|u|G_{ab}(\mathbf{i})^2-\frac{2}{|u|+3|v|}\big(|u|T_a(\mathbf{i})+\sqrt{3}|v|V'_a(\mathbf{i})\big)^2,\\
H_{G_2,\text{III}}&=&\sum_\mathbf{i} -|u|M_{ab}(\mathbf{i})M_{ab}(\mathbf{i})-|v|M'_{ab}(\mathbf{i})M'_{ab}(\mathbf{i})\nn\\
&=&\sum_\mathbf{i} -\big(|u|+\frac{1}{2}|v|\big)G_{ab}(\mathbf{i})^2-2(|u|-|v|)T_a(\mathbf{i})^2,\\
&=&\sum_\mathbf{i} -\big(\frac{1}{2}|u|+|v|\big)G_{ab}(\mathbf{i})^2-2(-|u|+|v|)T'_a(\mathbf{i})^2,\\
H_{G_2,\text{IV}}&=&\sum_\mathbf{i} -2|u|V_a(\mathbf{i})V_a(\mathbf{i})-|v|M'_{ab}(\mathbf{i})M'_{ab}(\mathbf{i})\nn\\
&=&\sum_\mathbf{i} -|v|G_{ab}(\mathbf{i})^2-\frac{2}{3|u|+|v|}\big(\sqrt{3}|u|V_a(\mathbf{i})-|v|T'_a(\mathbf{i})\big)^2,\label{eq8}
\eeq
where Roman number subscripts represent quadrants. Here we use $H_{G_2,\text{II}}$ as an example to show the derivation of these decompositions. Mean-field decompositions in other quadrants can be similiarly deduced.
\beq
H_{G_2,\text{II}}&=&u \sum_\mathbf{i} C_A(\mathbf{i}) +v \sum_\mathbf{i} C_B(\mathbf{i})\nn\\
&=&-|u|\sum_\mathbf{i} M_{ab}(\mathbf{i})M_{ab}(\mathbf{i}) +|v| \sum_\mathbf{i} M'_{ab}(\mathbf{i})M'_{ab}(\mathbf{i}).
\eeq
By means of Eq.~(\ref{eq:54}) (Fierz identity),
\beq
H_{G_2,\text{II}}&=&\sum_\mathbf{i} -|u|M_{ab}(\mathbf{i})M_{ab}(\mathbf{i}) -|v|V'_{a}(\mathbf{i})V'_{a}(\mathbf{i}).
\eeq
By means of Eq.~(\ref{eq:52}),
\beq
H_{G_2,\text{II}}&=&\sum_\mathbf{i} -|u|G_{ab}(\mathbf{i})G_{ab}(\mathbf{i})-2|u|T_a(\mathbf{i})T_a(\mathbf{i}) -|v| V'_{a}(\mathbf{i})V'_{a}(\mathbf{i}).
\eeq
By means of Eq.~(\ref{eq:53}) and $V_a=\sqrt{3}T_a-2V'_a$, we have $T_aV'_a+V'_aT_a=\sqrt{3}T_aT_a+\frac{1}{\sqrt{3}}V'_aV'_a$, and hence
\beq
H_{G_2,\text{II}}&=&\sum_\mathbf{i} -|u|G_{ab}(\mathbf{i})^2-\frac{2}{|u|+3|v|}\big(|u|T_a(\mathbf{i})+\sqrt{3}|v|V'_a(\mathbf{i})\big)^2.
\eeq
The order parameters in the first quadrant form a $G_{2}$ vector $\sum_\mathbf{i} \text{sgn}(\mathbf{i})(|u|V_a(\mathbf{i})+|v|V'_a(\mathbf{i}))$ and are the mixtures of $T_a$ and $V_a$, which is consistent with the analysis in the main text. Situations are more complicated in the second, third and fourth quadrants, where the order parameters are generally speaking $G_2$ generators $\sum_\mathbf{i} \text{sgn}(\mathbf{i})G_{ab}(\mathbf{i})$ and $G_2$ vectors.

When the $G_2$ symmetry is spontaneously broken to $SU(3)$, the order parameter can be chosen as (take $H_{G_2,\text{I}}$ and the 4th direction as an example)
\beq
O_4&=&\sum_\mathbf{i}\text{sgn}(\mathbf{i})\big(|u|V_4(\mathbf{i})-|v|V^\prime_4(\mathbf{i})\big)\nn\\
&\sim &\sum_\mathbf{i}\text{sgn}(\mathbf{i})\big(\cos\theta V_4(\mathbf{i})-\sin \theta T_4(\mathbf{i})\big),
\eeq
with $0\le \theta\le \frac{\pi}{3}$, where $\theta=0$ and $\frac{\pi}{3}$ correspond to $v=0$ and $u=0$, respectively. $O_4$ is diagonal in Nambu basis, and the $SU(3)$ generators commuting with $O_4$ can be constructed as
\beq
&&\Lambda_1(\mathbf{i})=-G_{12}(\mathbf{i})-G_{35}(\mathbf{i}),~\Lambda_2(\mathbf{i})=-G_{46}(\mathbf{i})-2G_{25}(\mathbf{i}),~\Lambda_3(\mathbf{i})=G_{15}(\mathbf{i})+G_{23}(\mathbf{i}),~\Lambda_4(\mathbf{i})=G_{45}(\mathbf{i})-2G_{26}(\mathbf{i}),\nn\\
&&\Lambda_5(\mathbf{i})=-G_{14}(\mathbf{i})+2G_{36}(\mathbf{i}),~\Lambda_6(\mathbf{i})=G_{24}(\mathbf{i})-2G_{56}(\mathbf{i}),~\Lambda_7(\mathbf{i})=G_{34}(\mathbf{i})+2G_{16}(\mathbf{i}),~\Lambda_8(\mathbf{i})=\sqrt{3}G_{06}(\mathbf{i}),
\eeq
which satisfy the standard commutation relations of $SU(3)$ Gell-Mann matrices. Note that $\Lambda_3(\mathbf{i})$ and $\Lambda_8(\mathbf{i})$ are diagonal and form the same Cartan subalgebra as $G_2$. For the order parameter rotated from $O_4(\mathbf{i})$ by $U(\mathbf{i})$, the corresponding $SU(3)$ generators become $\Lambda_a'(\mathbf{i})=U(\mathbf{i})\Lambda_a(\mathbf{i})U(\mathbf{i})^\dag$. When the $SU(3)$ symmetry is further broken into $SU(2)\times U(1)$, the remaining generators commute with both $O_4$ and $O^G_{06}=\sum_\mathbf{i}\text{sgn}(\mathbf{i})G_{06}(\mathbf{i})$ are $\Lambda_1(\mathbf{i}),~\Lambda_2(\mathbf{i}),~\Lambda_3(\mathbf{i}),~\Lambda_8(\mathbf{i})$.

\end{document}